\documentstyle[12pt,fleqn]{article}
\begin{document}
\setlength{\unitlength}{1mm}
\title{{\hfill {\small Alberta-Thy 33-95} } \vspace*{2cm} \\
Black Hole Entropy: Off-Shell vs On-Shell}
\author{\\
V.P.Frolov\footnote{e-mail: frolov@phys.ualberta.ca}${}^{1,2,3}$,
D.V.
Fursaev\footnote{e-mail: dfursaev@phys.ualberta.ca}${}^{1,4}$, and
A.I.Zelnikov\footnote{e-mail: zelnikov@phys.ualberta.ca}${}^{1,3}$
\date{}}
\maketitle
\noindent
$^{1}${ \em
Theoretical Physics Institute, Department of Physics, \ University of
Alberta, \\ Edmonton, Canada T6G 2J1}
\\ $^{2}${\em CIAR Cosmology Program}
\\ $^{3}${\em P.N.Lebedev Physics Institute,  Leninskii Prospect 53,
Moscow
117924, Russia}
\\ $^{4}${\em Bogoliubov Laboratory of Theoretical Physics, Joint
Institute for
Nuclear Research, \hfill \\ 141 980 Dubna, Russia}

\bigskip

\begin{abstract}
Different methods of calculation of quantum corrections to the
thermodynamical
characteristics of a black hole are discussed and compared. The
relation
between on-shell  and off-shell approaches is established. The
off-shell
methods are used to explicitly demonstrate that the thermodynamical
entropy
$S^{TD}$ of a black hole, defined by the first thermodynamical law,
differs
from the statistical-mechanical entropy $S^{SM}$, determined as
$S^{SM}=-\mbox{Tr}(\hat{\rho}^H\ln\hat{\rho}^H)$ for the density
matrix
$\hat{\rho}^H$  of a black hole. It is shown that the observable
thermodynamical black hole entropy can be presented in the form
$S^{TD}=\pi
{\bar r}_+^2+S^{SM}-S^{SM}_{Rindler}$. Here
${\bar r}_+$ is the radius of the horizon shifted because of the
quantum
backreaction effect, and $S^{SM}_{Rindler}$ is the
statistical-mechanical
entropy calculated in the Rindler space.
\end{abstract}

\bigskip

{\it PACS number(s): 04.60.+n, 12.25.+e, 97.60.Lf, 11.10.Gh}

\newpage
\baselineskip=.6cm

\noindent
\section{Introduction}
According to the  thermodynamical analogy in  black hole physics,
the
entropy of a black hole  in the Einstein theory of gravity is
 \begin{equation}\label{0.1}
{S}^{BH} ={A}^H
/(4l_{\mbox{\scriptsize{P}}}^2),
\end{equation}
where   ${A}^H$   is   the   area   of   a   black   hole   surface
and
$l_{\,\mbox{\scriptsize{P}}}=
(\hbar G/c^3)^{1/2}$   is   the   Planck length
\cite{Beke:72,Beke:73,Beke:74,Hawk:75}.
In  black hole physics the Bekenstein-Hawking entropy ${S}^{BH}$
plays
basically the same role as in the usual thermodynamics.  It can be
determined
by  the response of the free energy of a system containing a black
hole to the
change of the temperature of the system.

In the Euclidean approach
\cite{GiHa:76,Hawk:79,York:86,York:etal,BrBrWhYo:90}
the free
energy $F$ is directly related to  the Euclidean action  calculated
for the
regular Euclidean solution of the vacuum Einstein equations (the
Gibbons-Hawking instanton).  According to the first law of
thermodynamics the
thermodynamical entropy of a black hole $S^{TD}$ is defined  by the
relation
\begin{equation}\label{0.2}
dF=-S^{TD}dT~~~,
\end{equation}
where $T$ is the temperature of the system containing  a black hole.
The free energy $F$ besides the classical (tree-level) contribution
includes
quantum (one-loop) corrections. For this reason the thermodynamical
entropy in
addition to the classical (tree-level) part $S^{BH}$  acquires also a
quantum
correction $S^{TD}_1$
\begin{equation}\label{0.3}
S^{TD}=S^{BH}+S^{TD}_1.
\end{equation}
To find $S^{TD}$ one must compare two equilibrium configurations.
That is why
all the calculations which are required to determine $S^{TD}$ can be
made by
using the regular Gibbons-Hawking instanton as the background metric.
One
usually refers to these type of calculations  as to the {\em
on-shell} method.

The fundamental problem of black hole thermodynamics is its
statistical-mechanical foundation. The problem consists of the
following three
parts: (1) a definition of internal degrees of freedom of a black
hole; (2) the
calculation of the statistical-mechanical entropy $S^{SM}$ of a black
hole
$S^{SM}=-\mbox{Tr}(\hat{\rho}^H\ln \hat{\rho}^H )$  by counting the
dynamical
degrees of freedom described by the black hole density matrix
$\hat{\rho}^H$;
and (3) the establishing the relation between the
statistical-mechanical
$S^{SM}$ and the thermodynamical $S^{TD}$ entropies.

One of the ideas which was proposed is to identify the internal
degrees of
freedom of a black hole with its quantum excitations.
This idea has different realizations (see e.g.
Ref.\cite{Bekenstein,Frol:95a}
and references therein) and it has been widely discussed recently.
There is
enormous number of papers, where the statistical-mechanical entropy
has been calculated for different black hole models. The main purpose
of our
paper is to establish
the relation between the results of these calculations and the
observable
thermodynamical black hole entropy $S^{TD}$.

It should be stressed that the problem of relations between  $S^{TD}$
and
$S^{SM}$ is very nontrivial for black holes.
The quantities $S^{TD}$ and $S^{SM}$ are equal for the usual
thermodynamical
systems. Black holes  possess a property which singles them out of
the other
thermodynamical systems. Namely, in a state of thermal equilibrium a
mass $m$
of a black hole is a universal function of a temperature $T$. But the
mass
uniquely determines the geometry of a black hole, and hence the
internal
parameters of the Hamiltonian describing its quantum excitations.
This property
has two important consequences: (i)  $S^{TD}$ and $S^{SM}$ do not
coincide for
a black hole \cite{Frol:95}; (ii) Calculation of $S^{SM}$ and its
comparison
with $S^{TD}$ require {\em off-shell} methods. The latter means that
one needs
to consider the temperature $T$ and the mass of a black hole $m$ as
independent
parameters. The problem which arises is that when $T\neq T^{BH}\equiv
(8\pi
m)^{-1}$ there is no  regular complete vacuum Euclidean solutions.
For this
reason it is necessary either to consider the background metric which
is not a
solution of the vacuum gravitational equations, or to exclude some
region of
spacetime near the horizon and to make a solution incomplete. In both
cases the
calculation of the free energy meets problems. Moreover  the result
may depend
on the chosen concrete off-shell procedure\cite{fn1}.

In this paper we obtain the relation between different definitions of
the black
hole entropy. We also discuss and compare different off-shell methods
(brick
wall, conical singularity, blunt cone,
and volume cut-off), and their relations to the on-shell approach.
We
illustrate these relations for a simplified two-dimensional model,
where all
the calculations can be performed exactly.  It is  explicitly
demonstrated
that the thermodynamical entropy $S^{TD}$ of a black hole,  differs
from the
statistical-mechanical entropy $S^{SM}$. One of the main results is
the
observation that the one-loop contribution $S^{TD}_1$ of a quantum
field to the
thermodynamical entropy can be presented in the form
	\begin{equation}\label{01}
	S^{TD}_1=S^{SM}-S^{SM}_{Rindler}+\Delta S~~~.
	\end{equation}
Here $S^{SM}_{Rindler}$ is the statistical-mechanical entropy
calculated in the
Rindler space, and $\Delta S$ is an additional finite correction
caused by the
shift of the black hole horizon because of quantum effects. The
entropy
calculated using the brick-wall and volume cut-off methods is
directly related
with $S^{SM}$. This quantity is divergent (in 2D case) as
$\ln\epsilon$, where
$\epsilon$ is the proper distance to the horizon. On the other hand,
the
entropy calculated using the conical singularity and blunt cone
methods
coincides with the difference  $S^{SM}-S^{SM}_{Rindler}$.
It is finite because logarithmical divergence in $S^{SM}$ is exactly
canceled
by the  divergence of the Rindler entropy $S^{SM}_{Rindler}$.

It is well known that one-loop effective action which defines the
free energy
contains local ultraviolet divergences. In order to work with well
defined
finite quantities  it is necessary to renormalize it. Usually one
assumes that
the bare classical action contains the same local structures, that
arise in the
one-loop calculations. In the procedure of the renormalization one
excludes the
local one-loop divergences by a simple redefinition of  coupling
constants of
the classical action. In our approach we assume that this
renormalization
procedure has been done from the very beginning. We use  renormalized
observable quantities as  parameters of  on-shell solutions. In this
case  the
renormalized one-loop effective action is finite (at least on shell).
Quantum
effects which change this solution  can be considered as small
perturbations
for black holes with mass much larger than the Planckian mass. This
also allows
us to restrict ourselves by considering only those off-shell
solutions which
are close to the renormalized on-shell one\cite{fn2}.
As a result of our analysis we find out that all  thermodynamical
characteristics of a black hole expressed in terms of observable
parameters are
finite and their  definition does not require the knowledge of
physics at
Planckian scales.

The paper is organized as follows. In Section 2 we remind the main
features of
the Euclidean approach and give the general definition of the
thermodynamical
entropy which is used throughout this paper.
The description of a two-dimensional model is given in the Section 3.
This
Section also contains  the derivation of the on-shell free energy and
the
thermodynamical entropy for this model. The general scheme of the
off-shell
methods is discussed in Section 4. The off-shell effective action,
free energy,
and statistical-mechanical entropy are exactly calculated for four
the most
common off-shell approaches: brick wall (Section 5), conical
singularity
(Section 6), blunt cone (Section 7), and volume cut-off (Section 8)
methods.
Section 9 includes the comparison of the off-shell expressions for
free energy
and entropy, as well as  the relation between statistical-mechanical
and
thermodynamical entropies of a black hole.  Section 10 contains
concluding
remarks.
Important results concerning conformal transformations of the
effective action
in the presence of conical singularities, derivation of the effective
action on
a cylinder and the role of the vacuum polarization effect in the
brick wall
model, which are used in the main  text, are collected in the
Appendices.

\section{Euclidean Approach and Thermodynamical Entropy}
\setcounter{equation}{0}
The starting point of the Euclidean approach to the black hole
thermodynamics
is the partition function $Z(\beta)$ and the effective action
$W(\beta)$ which
for a canonical ensemble in the presence of black holes are defined
by the path
integral
	\begin{equation}\label{1.1}
	e^{-W(\beta)}=Z(\beta)=\int [D\phi] e^{-I[\phi]}.
	\end{equation}
Here $I[\phi]$ is the Euclidean classical action and all the physical
variables
$\phi$, including the gravitational field $g_{\mu\nu}$, are assumed
to be
periodic or antiperiodic, depending on their statistics, in the
Euclidean time
$\tau$ with the period $\beta_{\infty}$. As usual, the class of
metrics
involved in (\ref{1.1}) is supposed to be asymptotically flat . The
parameter $\beta_{\infty}$ has the meaning of the inverse temperature
measured
at
the spatial infinity. It is also assumed that the integration measure
$[D\phi]$
is defined as the covariant measure.

The standard way to calculate $W$ is to use quasiclassical
approximation.
Thus, if $\phi_0$ is a stationary point of $I[\phi]$
\begin{equation}\label{1.2}
\left. {\delta I\over{\delta \phi}}\right| _{\phi=\phi_0} =0~~~,
\end{equation}
then one has the decomposition
\begin{equation}\label{1.3}
I[\phi_0 +\tilde{\phi}]=I[\phi_0]+I_2[\tilde{\phi}]+\ldots,
\end{equation}
where $I_2$ is a quadratic in fluctuations $\tilde{\phi}$ part of the
linearized action and the dots in the right-hand side denote the
terms of the
higher order in $\tilde{\phi}$.
Using this relation one gets
	\begin{equation}\label{1.5}
	Z(\beta)= e^{-I[\phi_0]}
	\int [D\tilde{\phi}] e^{-I_2[\tilde{\phi}]}\equiv
e^{-I[\phi_0]} Z_1(\beta)~ .
	\end{equation}
The result of the Gaussian integration over $\tilde{\phi}$ in
(\ref{1.5}) can
be  expressed in terms of the determinants of the
corresponding wave operators $D_j$ for the different spins $j$
	\begin{eqnarray}
	Z_1(\beta)\equiv Z_1 [\phi_0(\beta)]=\prod_{j}\{\det [ -\mu^2
D_j(\phi_0)]
\}^{ \mp{1 \over 2}}~~~. 	\label{1.6}
	\end{eqnarray}
Operators $D_j$ are determined by the quadratic part
$I_2 = {1 \over 2} \int d x \sqrt{g} \tilde{\phi} D_0 \tilde{\phi}$
of the
action and their explicit form depends on the spin $j$. For instance,
for the
conformally invariant massless scalar field in $d$ dimensional space
$D_0=\triangle-(d-2)(4(d-1))^{-1}R$, where
$\triangle=\nabla_\mu\nabla^\mu$ is
the Laplace operator and $R$ is the scalar curvature. A constant
$\mu^2$ in
(\ref{1.6}) is an
arbitrary renormalization parameter with the dimension of the length.
It does
not depend on the field configuration $\phi$.
Equation (\ref{1.6}) enables one to represent the effective action in
the
one-loop approximation as the sum
	\begin{equation}\label{1.4}
	W(\beta)=I[\phi_0(\beta)]-\ln Z_1(\beta)\equiv
I[\phi_0(\beta)]+W_1[\phi_0(\beta)] .
	\end{equation}
The one-loop contribution\cite{fn3}
$W_1[\phi_0]$ to the effective action is ultraviolet divergent and,
as usual,
the classical action $I$  is assumed
to be chosen in such a way that the corresponding local divergences
of $W_1$
can be removed by simple redefinition of the coupling constants in
$I$. From
now on we suppose that it has been done and that the classical action
is
written in terms of  renormalized coefficients, $\phi_0$ is its
extremum, and
$W_1$ is the {\it renormalized} one-loop action\cite{fn4}.
The ambiguity in the choice of the parameter $\mu$ in (\ref{1.6})
corresponds
to a freedom in the choice of {\it finite} counterterms which can be
added to
the action after renormalization.

To apply this general scheme to a black hole we assume that it is
non-rotating,
uncharged, and that there is no spontaneous symmetry breaking, so
that  average
values of all  fields except the gravitational one vanish. Also it is
worth taking the renormalized cosmological constant to be zero to
provide an
asymptotically flat black hole solution $g_0$
of the (vacuum) gravitational equations.  The solution
represents a Gibbons-Hawking instanton which is
regular at the Euclidean horizon. For the Einstein theory such an
instanton is
described by the Schwarzschild metric and depends only on one
constant -- mass
$m$ of a black hole. The condition of regularity of this metric at
the horizon
implies that $\beta_{\infty}=\beta_H=8\pi m$.

When  considering quantum corrections it is worth keeping in mind
that a system
for a chosen boundary conditions (periodicity in $\tau$) necessarily
consists
of a black hole in thermal equilibrium with a surrounding thermal
radiation
which also contributes into observable thermodynamical quantities.
This
contribution is infinite for the thermal bath of the infinite size.
Moreover,
an equilibrium of a black hole with an infinite bath is  unstable.
For this
reason it is important from the very beginning to consider a black
hole
surrounded by a boundary surface $B$ of a finite size
\cite{York:86,York:etal,BrBrWhYo:90}. We assume this surface cannot
be
penetrated by fields. This is provided by the corresponding boundary
conditions
on it. For simplicity $B$ is assumed to be spherical of a radius
$r_B$ and a
hole to be located in the center. For the Schwarzschild black hole
thermal
stability is guaranteed if $r_B<3m$. Finally, in such a formulation
of the
problem the parameter $\beta$ is the inverse temperature measured on
$B$.
Further we suppose
that all the necessary requirements of this kind are satisfied
and we omit their discussion.

Eq.(\ref{1.4}) contains the renormalized effective action $W$
calculated on a particular classical solution. This renormalized
action itself
is defined as a functional
\begin{equation}\label{1.51}
W[\phi]=I[\phi]+W_1[\phi]
\end{equation}
for an arbitrary field $\phi$ with appropriately chosen boundary
conditions.
The extremum $\bar{\phi}$ of this functional
\begin{equation}\label{1.7}
\left.
{\delta W\over{\delta \phi}}\right| _{\phi=\bar{\phi}} =0
\end{equation}
describes a modified field configuration which differs from a
classical solution by quantum corrections:
$\bar{\phi}=\phi_0 +\hbar\phi_1$.
The important observation is that, if one is interested in the
one-loop effects, the difference between the values of $W$ on
$\phi_0$ and
$\bar{\phi}$ turns out to be of the second order
in the Planck constant $\hbar$
\begin{equation}\label{1.71}
W(\beta)=W[\phi_0(\beta)]=W[\bar{\phi}(\beta)]+ O(\hbar^2)~~~.
\end{equation}
This follows from (\ref{1.7}), provided the quantum corrected and
classical
solutions obey the same boundary conditions.

The {\it thermodynamical entropy}  of a black hole $S^{TD}$ is
defined by  the
response of the  free energy $F(\beta) =\beta^{-1}W(\beta)$ to the
change of
the inverse temperature $\beta$ for fixed $r_b$.
\begin{equation}\label{1.72}
S^{TD}(\beta)=\beta^{2}  {d F(\beta)\over d\beta} =\left(\beta {d
\over
d\beta}-1\right)W(\beta)~~~.
\end{equation}
We remind that the renormalized effective action $W(\beta)$ is
calculated
on-shell, that is for $\beta_\infty=8\pi m$.
The thermodynamical entropy $S^{TD}$ can be written as
\begin{equation}\label{1.73}
S^{TD}=S_0^{TD}+S_1^{TD}~~~.
\end{equation}
It can be shown \cite{York:86,BrBrWhYo:90} that
\begin{equation}\label{1.73a}
S_0^{TD}=\left(\beta {d \over d\beta}-1\right)I[\phi_0(\beta)]~~~,
\end{equation}
coincides with the Bekenstein-Hawking entropy $S^{BH}$ given by
Eq.(\ref{0.1}),
while
\begin{equation}\label{1.74}
S^{TD}_1(\beta)=\left(\beta {d \over
d\beta}-1\right)W_1[\phi_0(\beta)]~
\end{equation}
describes the quantum correction to it. This correction contains also
the
entropy of the thermal radiation outside the black hole as its part.
By its
construction the thermodynamical entropy $S^{TD}$  is well defined
and finite.
All the calculations required to obtain this quantity can be
performed {\em
on-shell}, that is on a regular complete vacuum Euclidean solution of
the
gravitational equations. The parameters of this solution are
expressed only in
terms of the renormalized coupling constants.

\section{Description of the Model. On-Shell Results}
\setcounter{equation}{0}
In four dimensions the  calculation of $S^{TD}_1$ is a quite
complicated
problem. To discuss the properties of $S^{TD}_1$ and its relation to
$S^{SM}$
it is instructive to consider a simplified two-dimensional model
where the
calculations can be done explicitly. Certainly, the explicit forms of
these
quantities in two and in four dimensions are different. Nevertheless,
the study
of 2-D model allows
us to make definite conclusions concerning the physically interesting
case of a
four-dimensional spacetime. To preserve the maximal similarity with
the
four-dimensional case we consider a 2-D dilaton gravity  described by
the
following action
\begin{equation}\label{2.3}
I=-{1 \over 4} \int_{M^2}^{}(r^2 R+2(\nabla r)^2+2)\sqrt{\gamma}d^2x
 -{1 \over 2}\int_{\partial M^2}^{}r^2 (k-k_0)dy +\frac 12 \int
\sqrt{\gamma}\varphi_{,\mu}\varphi^{,\mu}.
\end{equation}
The 2-D metric $\gamma$, dilaton field $r$, and a scalar field
$\varphi$ are
dynamical variables of the problem. We denote by $R$ the curvature of
$\gamma$, and by $k$  the extrinsic curvature of $\partial M^2$. This
model is
similar to the one which has been extensively studied\cite{CGHS} as
an example
of a renormalizable exactly solvable theory of two-dimensional
dilaton gravity
coupled to matter.
In the absence of the scalar field $\varphi$ this action can be
obtained from
the 4-D Euclidean Einstein action
\begin{equation}\label{2.1}
I^{(4)}=-{1 \over 16\pi } \int_{M^4}^{}R^{(4)}\sqrt{g}d^4x-
{1 \over 8\pi }\int_{\partial M^4}^{}(K^{(4)}-K^{(4)}_0)\sqrt{h}d^3x
,
\end{equation}
by its reduction to the spherically symmetric metrics of the form
\begin{equation}\label{2.2}
ds^2=\gamma_{ab}dx^a dx^b+r^2d\omega^2 .
\end{equation}
Here $\gamma_{ab}$ is a 2-D metric,  $r$ is a scalar function on the
two-dimensional manifold, and $d\omega^2$ is the line element on the
unit
sphere. $K^{(4)}_0$ is the standard subtraction term, and
$k_0=K^{(4)}_0$.

Since the 2-D action $I$ is related with 4-D action $I^{(4)}$ by the
reduction
procedure, the pair of fields $(\gamma_0,\varphi_0)$, where
$\varphi_0=0$ and
$\gamma_0$ is a 2-D Schwarzschild metric
\begin{equation}\label{2.4}
ds^2=fd\tau^2+f^{-1}dr^2,\hspace{1cm}f=1-r_+/r ,
\end{equation}
is evidently the extremum of the functional $I$. The regularity
condition at
$r=r_+$ requires $\tau$ to be periodic with the period $\beta_H=4\pi
r_+$. The
Gibbons-Hawking instanton, i.e. the regular complete Euclidean
manifold with
the metric (\ref{2.4}), is shown in Fig.1.
\begin{figure}
\label{f1}
\let\picnaturalsize=N
\def\picsize{3cm}
\def\picfilename{fig1.eps}
\ifx\nopictures Y\else{\ifx\epsfloaded Y\else\input epsf \fi
\let\epsfloaded=Y
\centerline{\ifx\picnaturalsize N\epsfxsize \picsize\fi
\epsfbox{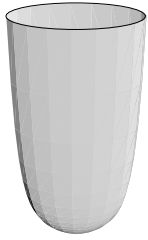}}}\fi
\caption[f1]{Embedding diagram for a two-dimensional  Gibbons-Hawking
instanton. Regularity condition at the Euclidean horizon $r=r_+$
requires
$\beta_\infty=\beta_H\equiv 8\pi m$. }
\end{figure}

Consider a region $M_B$ of the Gibbons-Hawking instanton within the
external
boundary $\Sigma_B$ at $r=r_B$ (see Fig.2). If the boundary
conditions are
fixed on the surface $\Sigma_B$, and $\beta$ is the proper length of
the line
$r=r_B$, then the classical Euclidean action calculated for the
region $M_B$
and expressed in terms of the boundary conditions ($\beta ,r_B$) is
\begin{equation}\label{2.5}
I(\beta, r_B)=I[\gamma_0,\varphi_0]=3\pi r_+^2-4\pi r_+r_B+\beta
r_B~~~,
\end{equation}
where $r_+$ is defined by the equation
\begin{equation}\label{2.5a}
\beta=4\pi r_+(1-r_+/r_B)^{1/2}~~~ ,
\end{equation}
and $\beta$ is the inverse temperature at $r=r_B$. In the limit
$r_B\rightarrow\infty$,
when $\beta=4\pi r_+$,
the classical action takes the simple form
\begin{equation}\label{2.51}
I(\beta)={1 \over 16\pi}\beta^2 .
\end{equation}

\begin{figure}
\label{f2}
\let\picnaturalsize=N
\def\picsize{5cm}
\def\picfilename{fig2.eps}
\ifx\nopictures Y\else{\ifx\epsfloaded Y\else\input epsf \fi
\let\epsfloaded=Y
\centerline{\ifx\picnaturalsize N\epsfxsize \picsize\fi
\epsfbox{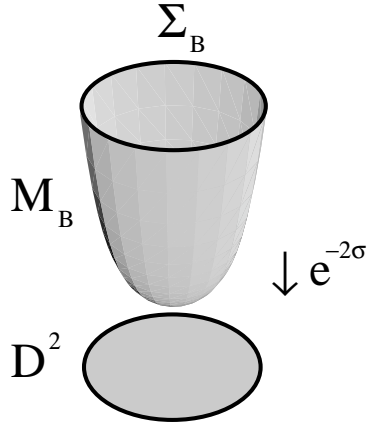}}}\fi
\caption[f2]{A region $M_B$ of the Gibbons-Hawking instanton with the
external
boundary $\Sigma_B$ at $r=r_B$. This region is conformal to the 2-D
flat unit
disk $D^2$.}
\end{figure}

In accordance with the general discussion of Section 2, the one-loop
contribution to the effective action is
	\begin{equation}\label{2.7}
	W_1(\beta)=\frac 12 \ln \det (-\mu^2 \Delta) .
	\end{equation}
Here the renormalized determinant is taken for the region $M_B$ of
the 2-D
instanton (\ref{2.4}). To make discussion more concrete we assume
that the
field $\varphi$ obeys   the Dirichlet boundary condition at the
mirror-like
boundary $\Sigma_B$ surrounding the black hole. The divergent part
which has
been removed from the action is
	\begin{equation}\label{2.7a}
	W_1^{div}[M_B]= -{1 \over 8\pi\delta}
\int_{M_B}\sqrt{\gamma}d^2x
	+{\ln \delta \over  12}~\chi[M_B]~~~,
	\end{equation}
	\begin{equation}\label{2.7aa}
	\chi[M_B]={1 \over 4\pi}\left(\int_{M_{B}} Rd^2x +
	2\int_{\Sigma_B} k dy\right)~~~,
	\end{equation}
where $\delta$ is the parameter of the ultraviolet regularization and
$\chi[M_B]$ is the Euler characteristics of the Gibbons-Hawking
instanton
$M_B$, which is the same as for a disk $\chi[M_B]=1$. To remove the
volume
divergence $\sim \int_{M_B}$ one must to introduce in the bare
classical action
a cosmological constant $\lambda$, which we put after renormalization
to be
$-1/2$, see (\ref{2.3}).
Removing of the other divergence in (\ref{2.7a}) requires
introduction of the
additional term in (\ref{2.3}),
but because it is just a topological invariant it can be neglected.

Using the conformal transformation the one-loop effective action
$W_1(\beta)$
can be found explicitly. Note that metric (\ref{2.4}) can be
represented in the
form
	\begin{equation}\label{2.71}
	ds^2=\left(1-{r_+ \over r}\right)d\tau^2+\left(1-{r_+ \over
r}\right)^{-1}dr^2=
	e^{2\sigma}d\tilde{s}^2 ,
	\end{equation}
	\begin{equation}\label{2.71a}
	d\tilde{s}^2=\mu^2\left(x^2d\tilde{\tau}^2+dx^2\right) .
	\end{equation}
Here
	\begin{equation}\label{2.72}
	\tilde{\tau}={\tau \over 2r_+}~~,~~0\leq \tilde{\tau}\leq
2\pi
	~~,~~x=\left({r-r_+ \over r_B-r_+}\right)^{\frac 12}e^{r-r_B
	\over 2r_+}~~,~~0\leq x \leq 1~,
	\end{equation}
and the conformal factor $\sigma$ is defined as
	\begin{equation}\label{2.73}
	\sigma(r)=\frac 12\left[\ln\left({r_B-r_+ \over r }\right)+
	{r_B-r \over r_+} + 2\ln{\left({2r_+\over \mu}\right)}
\right]~.
	\end{equation}
In order to preserve the dimensionality we introduce the parameter
$\mu$ with
the dimension of length into the flat space metric (\ref{2.71a}). The
above
conformal transformation
	\begin{equation}\label{2.73a}
	\gamma_{\mu\nu}\rightarrow\tilde{\gamma}_{\mu\nu}=
	e^{-2\sigma}\gamma_{\mu\nu}
	\end{equation}
is a map
of the region $M_B$ onto the flat 2-D disk $D^2$ of the unit radius
(measured
in units of $\mu$), see Fig.2.  It will be shown that the physical
results do
not depend on the particular choice of $\mu$\cite{fn5}.

For a conformal field the transformation law of $W_1$ under
this map  can be obtained by an integration of a conformal anomaly.
The
corresponding formulas are collected in the Appendix A. Denote by $C$
the
renormalized one-loop  effective action for the unit disk $D^2$,
Eq.(\ref{2.71a}),
then using the relation (\ref{a10}) we get
	\begin{equation}\label{2.75}
	W_1(\beta,r_B)=\tilde{W}_1(\beta,y(\beta,r_B)),
	\end{equation}
where $y=r_+/r_B$ and
	\begin{equation}\label{2.75a}
	\tilde{W}_1(\beta,y)={ 1\over 48}
	\left[-\frac 2y +2 \ln y +17 - 2 y - 13  y^2 \right] - \frac
16
\ln{\beta\over 2\pi\mu} + C~~~.
	\end{equation}
The relations (\ref{2.75}) and (\ref{2.75a}) require some
explanations. First
of all, the one-loop effective action $W_1(\beta,r_B)$ besides the
inverse
temperature $\beta$ at the boundary also depends  on its 'radius'
$r_B$. For
given $\beta$ and $r_B$ the gravitational radius $r_+$ is defined by
the
relation (\ref{2.5a}). To simplify the expressions we use the
dimensionless
variable $y=r_+/r_B$  instead of $r_B$. The relation (\ref{2.5a})
implies that
this dimensional variable $y$ is the function of $\beta$ and $r_B$
defined by
the following implicit relation
	\begin{equation}\label{2.75b}
	y (1-y)^{1/2}={\beta \over {4\pi r_B}} .
	\end{equation}

The one-loop contributions to the free energy $F_1$ and to the
thermodynamical
entropy $S_1^{TD}$ are defined by the
formulas
	\begin{equation}\label{2.75c}
	F_1(\beta,r_B)=\beta^{-1}W_1(\beta,r_B),\hspace{1cm}S_1^{TD}
	=\beta\left. {\partial {W_1(\beta,r_B)}\over \partial
\beta}\right|
_{r_B} 	-W_1(\beta,r_B)~~~.
	\end{equation}
The derivative of $W_1$ can be expressed in terms of the partial
derivatives of $\tilde{W}_1$
	\begin{equation}\label{2.75d}
	\left. {\partial {W_1(\beta,r_B)}\over \partial \beta}\right|
_{r_B}=\left.
{\partial 	{\tilde{W}_1(\beta,y)}\over \partial \beta}\right|
_{y}+\left.
{\partial 	{\tilde{W}_1(\beta,y)}\over \partial y}\right|
_{\beta}\left.
{\partial {y}\over 	\partial \beta}\right| _{r_B}~~~ ,
	\end{equation}
where
	\begin{equation}\label{2.75e}
	\left. {\partial {y}\over \partial \beta}\right|
_{r_B}={2y(1-y)\over{\beta (
2-3y)}}~~~.
	\end{equation}
The latter equality results from Eq.(\ref{2.75b}).
Using the relations (\ref{2.75c})-(\ref{2.75e}) we finally obtain
	\begin{equation}\label{2.76}
	S_1^{TD}(y,\beta)={1 \over 48(2-3y)}\left[\frac 8y -13 y -28
y^2
	+13 y^3\right]-{1 \over 24} \ln y + \frac 16 \ln{\beta \over
2\pi\mu} -
	{17 \over 48}-C~.
	\end{equation}
This quantity is finite. The dimensionless constant $C$ does not
depend on the
parameters of the system and reflects the ambiguity in the definition
of the
entropy. For further consideration this ambiguity is not important,
so that
this and other similar constants can be omitted. For  a large value
of the
radius $r_B$ of the boundary  ($r_B\gg r_+$ or $y\ll 1$) the leading
term in
$S_1^{TD}$
is ${\pi \over 3} r_B\beta ^{-1}$. This leading term  coincides with
the
entropy of the one-dimensional thermal gas of massless scalar quanta.
It should
be noted that we always consider the case when $r_B<3/2r_+$, so that
the limit
discussed above has only formal meaning. The quantity $S_1^{TD}$ is
infinite
when $r_B=\frac 32 r_+$. This singularity also results in the
infinite heat
capacity at $y=3/2$. One can expect the same behavior of these
quantities in
four-dimensional case.

\section{Off-Shell Methods}
\setcounter{equation}{0}
In the above consideration we used the relation (\ref{2.5a}) which
can be
rewritten as $\beta_{\infty}=\beta_H$, where $\beta_{\infty}=\beta
(1-r_+/r_B)^{-1/2}$ denotes the inverse temperature on the boundary
$\Sigma_B$
as seen from infinity, and $(1-r_+/r_B)^{1/2}$ is the red-shift
factor.
$\beta_H$ is the inverse Hawking temperature (also measured at
infinity). The
relation $\beta_{\infty}=\beta_H$ has evident meaning of the
equilibrium
condition between the thermal radiation and the black hole and it is
this
relation which is assumed when we are speaking about the on-shell
quantities.

In the  next sections we consider  different off-shell approaches in
which the
condition $\beta_{\infty}=\beta_H$ is violated for the background
geometries.
The one-loop contribution to the effective action in these cases is
the
function of the three variables $\beta,~r_B,~r_+$:
$W^{\bullet}_1(\beta,~r_B,~r_+,\ldots)$. We use the superscript
$\bullet$ to
indicate that this quantity depends on the chosen off-shell
procedure. The dots
$\ldots$ in the argument of $W^{\bullet}_1$ indicate that it may also
depend on
some additional   parameters, which are different for different
off-shell
procedures. These parameters are not important now and will be
specified later.

In the general case the off-shell  entropy is defined by the response
of the
off-shell free energy $F^{\bullet}=\beta^{-1}W^{\bullet}$ on the
change of the
temperature, under the condition that the other parameters which
specify the
system ($r_B$) as well as the black hole ($r_+$) are fixed. According
to this
definition the one-loop off-shell entropy is
	\begin{equation}\label{entr1}
	S^{~\bullet}_1=\beta\left. {\partial {W_1^{~\bullet}}\over
\partial
\beta}\right|_{r_B,r_+,\ldots}-W_1^{~\bullet} .
	\end{equation}
It is assumed that the on-shell limit in (\ref{entr1}) is taken
at the end of the computation. This means that $r_+$ which enters
$S^{~\bullet}_1$ is put equal to its on-shell value, determined by
solving the
corresponding gravitational equations.

It occurs that the explicit formulas for $W^{\bullet}_1$ and
$S^{~\bullet}_1$
are greatly simplified if instead of $r_B$ and $r_+$ the following
dimensionless variables are used
	\begin{equation}\label{alpha}
	y=y(r_B,r_+)={r_+ \over
r_B}~~~,~~~\alpha=\alpha(\beta,r_B,r_+)
	={\beta_\infty \over \beta_H}={\beta \over 4\pi r_+
	\sqrt{1 - {r_+ \over r_B}}}~~~.
	\end{equation}
The variable $\alpha$ is the {\em off-shell parameter} so that the
condition
that a system is on-shell reads $\alpha=1$. The parameter $y$ is the
ratio of
the values of the dilaton field on the external boundary $\Sigma_B$
and on the
horizon.
We shall  use the notation
	\begin{equation}
	W_1(\beta,r_B,r_+,\ldots)=
\tilde{W}_1(\beta,\alpha(\beta,r_B,r_+), y(r_+,r_B), 	\ldots) .
	\end{equation}
For  fixed values of $r_+$ and $r_B$ the quantity $y=r_+/r_B$ is also
fixed,
while Eq.(\ref{alpha}) implies that $\alpha$  is proportional to
$\beta$. Thus
one has
	\begin{equation}\label{entr2}
	S^{~\bullet}_1=\beta\left.
	{\partial{\tilde{W}_1^{~\bullet}(\beta,\alpha,y,\ldots)}\over
\partial
\beta}\right|_{\alpha,y,\ldots}+\alpha\left. {\partial
{\tilde{W}_1^{~\bullet}(\beta,\alpha,y,\ldots)}\over \partial
\alpha}\right|_{\beta,y,\ldots}-W_1^{\bullet}~~~.
	\end{equation}
As earlier, it is assumed that after the calculations one must put
$\alpha=1$
in the right-hand-side of this relation. Then the corresponding
on-shell value
of $S^{~\bullet}_1$ depends only on the boundary conditions $\beta$
and $r_B$.
After these general remarks consider concrete off-shell methods.

\section{Brick wall model}
\setcounter{equation}{0}
\subsection{Effective action}
As the first example of the off-shell procedure we consider the so
called {\em
brick-wall model}, proposed by t'Hooft \cite{Hoof:85} and discussed
later in
many subsequent papers\cite{Mann,SU,Ghosh,KabStr,BE,Liberati,DLM}.
The basic
idea of this method is to introduce at some small proper distance
$\epsilon$
from the black hole horizon
an additional mirror-like boundary $\Sigma_{\epsilon}$. Denote by
$M_{B,\epsilon}$ the region located between $\Sigma_B$ and
$\Sigma_{\epsilon}$
(see Fig.3). To be more specific, assume, following to 't~Hooft, that
the field
$\varphi$ obeys  the Dirichlet condition on both boundaries
$\Sigma_B$ and
$\Sigma_{\epsilon}$.
The starting point of the brick-wall model is the partition function
$Z_1^{BW}(\beta)$ of massless scalar field in the region
$M_{B,\epsilon}$ near
the Schwarzschild black hole of the mass $m$
	\begin{equation}\label{4.0}
	\ln{Z_1^{BW}(\beta)}=-\frac 12 \ln\det(-\mu^2\triangle)~~~ .
	\end{equation}
Here $\beta$ is the inverse temperature  measured at $\Sigma_B$,
$"\ln\det"$ is
understood as renormalized quantity, and $\triangle$ is the Laplace
operator
for the scalar field in the region $M_{B,\epsilon}$ with the
Dirichlet boundary
conditions.  Because of the presence of the inner boundary
$\Sigma_{\epsilon}$
the region near the black hole horizon where the thermal gas cannot
penetrate
is completely excluded. For this reason the system is non-singular
for any
relation between the parameters $\beta$ and $m$, and the brick-wall
model can
be used for an off-shell extension.
To distinguish the quantities calculated in this off-shell procedure
we use
the abbreviation $BW$ as the superscript. The corresponding partition
function
$Z_1^{BW}$ and action $W_1^{BW}$ depend, in addition to $\beta$ and
$r_B$, on
$\epsilon$ and the value $r_+$ of the dilaton field on the horizon.
Our purpose
now is to find  $W_1^{BW}(\beta,r_B,r_+,\epsilon)$.

Obviously, this problem can be reduced to the calculation of the
effective
action  for some 'standard'  2-D flat region. We choose a cylinder as
such a
region   (see Fig.3).
\begin{figure}
\label{f3}
\let\picnaturalsize=N
\def\picsize{12cm}
\def\picfilename{fig3.eps}
\ifx\nopictures Y\else{\ifx\epsfloaded Y\else\input epsf \fi
\let\epsfloaded=Y
\centerline{\ifx\picnaturalsize N\epsfxsize \picsize\fi
\epsfbox{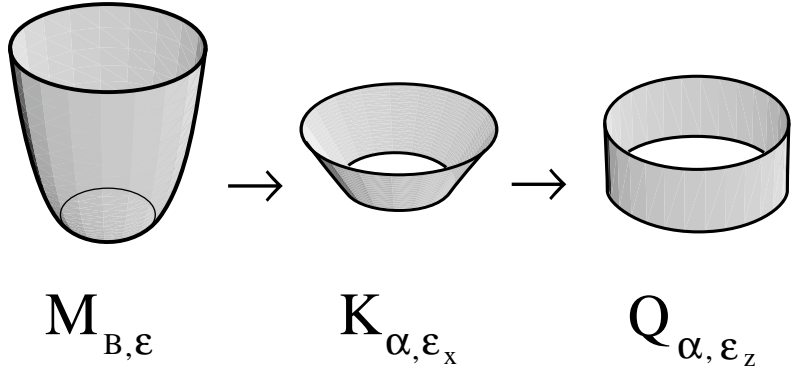}}}\fi
\caption[f3]{Conformal maps of the region $M_{B,\epsilon}$ of the
Gibbons-Hawking instanton onto the part  $K_{\alpha,\epsilon_x}$ of
the cone
$C_{\alpha}$, and of the region  $K_{\alpha,\epsilon_x}$ onto the
cylinder
$Q_{\alpha,\epsilon_z}$. $\epsilon$ is the proper distance of the
inner
boundary $\Sigma_{\epsilon}$ of $M_{B,\epsilon}$ to the horizon. The
parameter
$\epsilon_x$ is the distance from $\Sigma_B$ to the vertex of the
cone along
the cone generator,
and  $\epsilon_z$ is the length of the cylinder generator
(both measured in the units of $\mu$). The circumference length of
the
cylinder, as well as the  circumference length of of the external
boundary
$\Sigma_B$ of the cone, (measured in units $\mu$) is $2\pi\alpha$.}
\end{figure}

It is convenient to make the conformal transformation into two steps.

First, use the map (\ref{2.73a})  with $\sigma$ given by
Eq.(\ref{2.73}). Under
this transformation the metric takes the form
	\begin{equation}\label{4.3}
	d\tilde{s}^2=\mu^2(x^2d\tilde{\tau}^2+dx^2)~~,~~0\leq
\tilde{\tau}\leq  2\pi
\alpha~,~~\epsilon_x\leq x \leq 1 .
	\end{equation}
The embedding diagram for this space is shown in Fig.3. It is a part
$K_{\alpha,\epsilon_x}$ of the cone $C_{\alpha}$  between the
surfaces
$\Sigma_B$ located at $x=1$ and $\Sigma_{\epsilon}$ at $\epsilon_x$.
The value
of $x=\epsilon_x$ is related with the proper distance $\epsilon$ as
	\begin{equation}\label{dist}
	\epsilon_x=\epsilon~{2\pi \alpha \over \beta}\sqrt{y}
\exp{y-1 \over 2y}.
	\end{equation}
where the parameters $y$ and $\alpha$ are defined in
Eq.(\ref{alpha}).

Second, map $K_{\alpha,\epsilon_x}$ onto a cylinder
$Q_{\alpha,\epsilon_z}$
with the metric $\mu^2(d\tilde{\tau}^2+dz^2)$
	\begin{equation}\label{4.4}
	d\tilde{s}^2=\mu^2(x^2d\tilde{\tau}^2+ dx^2)=x^2
[\mu^2(d\tilde{\tau}^2 +
dz^2)]~,~~~z=\ln x~~~,
	\end{equation}
The cylinder has the circumference length $2\pi\alpha$ and the length
of its
generator is $\epsilon_z=-\ln \epsilon_x$ (in the $\mu$ units) (see
Fig.3).

Thus, the effective action $W_1^{BW}(\beta,r_B,r_+,\epsilon)$ can be
obtained
by conformal transformation, provided one knows the action
$W_1[Q_{\alpha,\epsilon_z}]$ for the 'standard' cylinder
$Q_{\alpha,\epsilon_z}$. It can be shown (see Appendix B) that
	\begin{equation}\label{4.5}
	W_1[Q_{\alpha,\epsilon_z}]=-\ln \mbox{Tr e}^{-2\pi\alpha
\mu\hat{H}}~~~,
	\end{equation}
where $\hat{H}$ is the Hamiltonian for the scalar massless field on
the
interval $(0,\mu\epsilon_z)$ with the Dirichlet boundary conditions
at the
ends.
Using this fact we get for $\epsilon_z\gg1$ (see Appendix B)
	\begin{equation}\label{4.6}
	W_1[Q_{\alpha,\epsilon_z}]=- {1 \over 12
\alpha}\epsilon_z-\frac 12
	\ln{\pi\alpha \over \epsilon_z} + o\left({1 \over
\epsilon_z}\right)~~~ .
	\end{equation}
The scale parameter $\mu$ disappears from  this expression because of
the
scale invariance of the action on the cylinder.
The effective action $W_1[K_{\alpha,\epsilon_x}]$ for the region
$K_{\alpha,\epsilon_x}$ obtained from $W_1[Q_{\alpha,\epsilon_z}]$ by
conformal
transformation has the form
	\begin{equation}\label{4.7}
	W_1[K_{\alpha,\epsilon_x}]=W_1[Q_{\alpha,\epsilon_z}]-
	{\alpha\over 12}\epsilon_z ~~~
	\end{equation}
while the transformation (\ref{2.73a}) gives
	\begin{equation}\label{4.7a}
	W_1[M_{B,\epsilon}]=W_1[K_{\alpha,\epsilon_x}]+\alpha
f(y)~~~,
	\end{equation}
	\begin{equation}\label{4.7aa}
	f(y)=-{1 \over 48}\left(-\frac 2y +2\ln y +2y
+13y^2-13\right)~~~.
	\end{equation}
The final result is obtained by using the formulas
(\ref{4.6})-(\ref{4.7a}).
The effective action \\$W_1^{BW}(\beta,r_B,r_+,\epsilon)$ written as
the function
of  $(\beta,\alpha,y,\epsilon)$ is
	\begin{equation}\label{4.8}
	W_1^{BW}(\beta,r_B,r_+,\epsilon)=
	\tilde{W}_1^{BW}(\beta,\alpha(\beta,r_B,r_+),y(r_B,r_+),
\epsilon)~~~ ,
	\end{equation}
	\begin{equation}\label{4.9}
	\tilde{W}_1^{BW}(\beta,\alpha,y,\epsilon)=
	{1 \over 12}\left(\alpha +{1 \over \alpha}\right)\ln{2\pi
\alpha\epsilon \over
\beta}-\frac 12\ln{\pi\alpha \over \ln{(\beta/2\pi\alpha\epsilon)}}
	\end{equation}
	\[
	+
	{\alpha \over 48}\left(15 -2y-13y^2\right)
	+{1 \over 24 \alpha}\left(1-\frac 1y +\ln
y\right)+o(\ln^{-1}(\beta/\epsilon))~~~.
	\]
For $\alpha=1$, i.e. on-shell this action can be represented as the
sum
	\begin{equation}\label{4.9a}
	\tilde{W}_1^{BW}(\beta,\alpha=1,y,\epsilon)=
	\tilde{W}_1(\beta,y)+\frac
16\ln\epsilon -\frac 12\ln{\pi \over
\ln{(\beta/2\pi\epsilon)}}+o(\ln^{-1}(\beta/\epsilon))
	\end{equation}
of the thermodynamical action $\tilde{W}_1(\beta,y)$ for the region
$M_B$ given
by Eq. (\ref{2.75a}) and an additional term which arises because of
the
presence of the wall. The latter diverges logarithmically in the
limit
$\epsilon\rightarrow 0~$
\cite{fn6}.

\subsection{Entropy}
The entropy $S_1^{BW}$ for the brick wall model is defined  by Eq.
(\ref{entr1}) using $W_1^{BW}$.
Written in terms of $(\beta,\alpha,y,\epsilon)$ it reads
	\begin{equation}\label{4.12}
	S^{BW}_1(\beta,\alpha,y,\epsilon )={1 \over 12\alpha}\left(
	2\ln{{\beta \over 2\pi\alpha \epsilon}}- \ln y + {1 \over
y}-1\right)
	\end{equation}
	\[
	+\frac 12 \ln{\pi\alpha \over \ln{(\beta/2\pi
\alpha\epsilon})}+
o(\ln^{-1}(\beta/\epsilon))~~~.
	\]
The on-shell value  of  $S^{BW}_1$ is obtained if one puts $\alpha=1$
in this
expression.

Note that the renormalization parameter $\mu$ does not enter
Eqs.(\ref{4.9})
and (\ref{4.12}), so that neither brick-wall action $W_1^{BW}$ nor
the entropy
$S_1^{BW}$ depends on it. It happens because under a constant
conformal
transformation the effective action acquires an addition proportional
to the
Euler characteristic of the manifold.
But the topology of $M_{B,\epsilon}$ is the topology of a cylinder
and its
Euler number is zero. Thus the effective action is invariant
under the constant  rescaling, and it does not depend on $\mu$.
On the other hand, the Euler
characteristic  of the complete regular instanton is the same as that
of the
disk $D^2$, and it does not vanish. As the result
the integral of the anomaly also does not vanish, and $\mu$ appears
in the
thermodynamical action and entropy
as a parameter of the dimensional transmutation.

	We show now that the brick wall entropy (\ref{4.12})
coincides with the
statistical mechanical entropy and can be represented in the form
	\begin{equation}\label{sm-entr}
	S_1^{BW}(\beta,\alpha,y,\epsilon)=
	-\mbox{Tr}\left(\hat{\rho}^H_\epsilon
	(\beta)\ln\hat{\rho}^H_\epsilon(\beta)\right)~~~.
	\end{equation}
Here $\hat{\rho}^H_\epsilon(\beta)$ is  the thermal density matrix
for the massless gas in the region $M_{B,\epsilon}$ near the black
hole,  $\beta$ being the inverse temperature measured at $\Sigma_B$.
In the t'Hooft's brick wall model this thermal gas is identified with
internal
degrees of freedom of the black hole.

To prove Eq.(\ref{sm-entr}) we obtain at first expression
(\ref{4.12}) for
$S^{BW}_1$ in a slightly different way. Eqs.(\ref{4.7}) and
(\ref{4.7a}) show
that
	\begin{equation}\label{4.15}
	W_1^{BW}(\beta,r_B,r_+,\epsilon)=\alpha f(y)-{\alpha
\epsilon_z \over
12}+W_1[Q_{\alpha,\epsilon_z}]~~~ .
	\end{equation}
To get $S^{BW}_1$ we keep the variables $r_B$, $r_+$, and $\epsilon$
fixed.
Under these conditions $y$ does not depend on $\beta$, while $\alpha$
is
proportional to $\beta$. As the result, the first two terms in
Eq.(\ref{4.15})
do not contribute into $S^{BW}_1$, so that
	\begin{equation}\label{4.16}
	S_1^{BW}=\left(\alpha{\partial \over \partial
\alpha}-1\right)
W_1[Q_{\alpha,\epsilon_z}]
	={1\over 6\alpha} \epsilon_z +\frac 12
	\ln{\pi\alpha \over \epsilon_z}+ o(\epsilon_z^{-1})~~~.
	\end{equation}
It can be easily verified that this expression coincides with
Eq.(\ref{4.12}).
Note  that $W_1[Q_{\alpha,\epsilon_z}]$ is given by (\ref{4.5}). The
quantity
$\left( 1-\beta{\partial \over \partial \beta}\right) \ln\mbox{Tr
e}^{-\beta
\hat{H_L}}$ can be identically rewritten as
$-\mbox{Tr}(\hat{\rho}_{L}(\beta)\ln\hat{\rho}_{L}(\beta))$, where
$\hat{H}_L$
is the Hamiltonian on the interval of the length $L$, and
$\hat{\rho}_{L}(\beta)=\rho_0\mbox{e}^{-\beta \hat{H_L}}$. Using this
relation
we can present (\ref{4.16}) in the form
	\begin{equation}\label{4.17}
	S_1^{BW}=-\mbox{Tr}(\hat{\rho}_{\mu\epsilon_z}
	(2\pi\mu\alpha)\ln\hat{\rho}_{\mu\epsilon_z}
        (2\pi\mu\alpha))~~~ .
	\end{equation}
This relation explicitly demonstrates that $S_1^{BW}$ is the entropy
of the
one-dimensional thermal gas on the interval $\mu\epsilon_z$ and with
the
temperature $(2\pi\mu\alpha)^{-1}$.
(The parameter $\mu$ is absent in (\ref{4.16}) for the reason
explained above.)

This result can be used to prove the formula (\ref{sm-entr}) because
the
density matrix $\hat{\rho}_{\mu\epsilon_z}(2\pi\mu\alpha)$ coincides
with the
black hole density matrix $\hat{\rho}^H_\epsilon(\beta)$. Indeed, we
used
conformal transformations which preserve the symmetry (a Killing
vector) and do
not affect the boundary conditions. Under these conditions the
Hamiltonian of
the conformal massless field is invariant, so that the density matrix
is also
invariant. Note however, that scales we used to define the
temperature and
distance may change. In order to define energy, temperature, etc., we
must fix
the normalization of the Killing vector. For the problem in question
we chose
the condition $(\xi^2)_{B}=1$ at the external boundary $\Sigma_B$. If
the
conformal factor $\sigma$ does not vanish on the boundary
($\sigma_B\ne 0$),
one must rescale $\xi^{\mu}\rightarrow \tilde{\xi}^{\mu}=\exp
(-\sigma_B)\xi^{\mu}$ to get  $\tilde{\xi}^2=1$ at the boundary after
the
conformal transformation. We have
	\begin{equation}\label{4.18}
	\mbox{e}^{-\beta \hat{H}_L}=\mbox{e}^{-\tilde{\beta}
\hat{\tilde{H}}_{\tilde{L}}} ,
	\end{equation}
where $\tilde{\beta}=\exp (-\sigma_B)\beta$, $\hat{\tilde{H}}=\exp
(\sigma_B)\hat{H}$, and $\tilde{L}$ is the proper length of the
interval in the
conformally related metric
$\tilde{\gamma}_{\mu\nu}=e^{-2\sigma}\gamma_{\mu\nu}$.

In particular for the conformal map (\ref{2.71}), (\ref{2.73}) which
we used as
the first step, Eq.(\ref{4.18}) implies
	\begin{equation}\label{4.19}
	\hat{\rho}^H_{\epsilon}(\beta)=\hat{\rho}^R_{\mu\epsilon_x}
	(2\pi\mu\alpha)~~~ .
	\end{equation}
Here $\hat{\rho}^H$ is the original  black-hole density matrix, and
$\hat{\rho}^R$ is a thermal density matrix in a Rindler space with
the metric
	\begin{equation}\label{4.20}
	d\tilde{s}^2=\mu^2[x^2d\tilde{\tau}^2+dx^2]=\left(
{X\over\mu}\right)^2dT^2+dX^2 .
	\end{equation}
The inverse temperature $2\pi\mu\alpha$ in the Rindler space is
measured at the
point of the boundary $X=\mu$, where the $g_{TT}=1$. The parameter
$\mu\epsilon_x$ is the proper distance from the inner boundary to the
horizon,
measured in the Rindler metric. Note, that the proper distance is not
conformal
invariant. Finally, by mapping Rindler space onto the flat one
(the corresponding transformation of the effective action from
$K_{\alpha,\epsilon_x}$ to $Q_{\alpha,\epsilon_z}$ is given by
(\ref{4.4})),
one receives the identity
	\begin{equation}\label{4.21a}
	\hat{\rho}^R_{\mu\epsilon_x}(2\pi\mu\alpha)=
\hat{\rho}_{\mu\epsilon_z}
	(2\pi\mu\alpha)~~~
	\end{equation}
between the Rindler density matrix and that on the interval. The
statistical-mechanical formula (\ref{sm-entr}) for $S_1^{BW}$
follows from the identities (\ref{4.17}), (\ref{4.19}) and
(\ref{4.21a}).

\section{Conical singularity method}
\setcounter{equation}{0}
Instead of excluding  the $\epsilon$-domain near the horizon, one can
work
directly on the complete
black hole geometry. However, if $\beta_\infty$ differs from the
Hawking value
$\beta_H$, the spacetime is not anymore  regular  because of
the presence of the conical singularity with the angle deficit
$2\pi(1-\alpha)$ at the horizon $r=r_+$ (fixed point of the Killing
vector).
Such a space has the $\delta$-like curvature located on the cone
vertex. For
this reason it is not a solution of the vacuum Einstein equations. We
call such
a space a {\em singular instanton} and denote it $M_B^\alpha$, see
Fig.4.

It is possible to develop the one-loop calculations working directly
on the
manifolds with this kind of singularities.  We refer to the
corresponding
approach as to the {\em conical singularity} method
\cite{SU,CalWil94,a3,F:MPL,SS1,F:PR,LW,Kabat,KSS,SS2}. The difference
between
it and quantum theory on the regular spaces is in the structure of
the
ultraviolet divergences \cite{DF:94(2),F:MPL,Dowker:9495}. Conical
singularities result in appearing in the effective action of
additional
divergent terms concentrated on the horizon surface and their
renormalization
requires new counterterms. The important
property however is that these counterterms turn out to be of the
order $(\beta_\infty-\beta_H)^2$ $\sim(1-\alpha)^2$ and hence when
taken
on-shell
they contribute neither to the entropy, nor to the free energy of the
black
hole \cite{a3,FS2,DLM,LW}.

In two dimensions, as follows from Eqs.(\ref{a3}),(\ref{a4}), the
divergent
part of the action on the singular instanton $M_B^\alpha$ can be
represented as
	\begin{equation}\label{divsing}
	W_1^{div}[M_B^\alpha]= -{1 \over 8\pi\delta}
\int_{M_B^\alpha}\sqrt{\gamma}d^2x
	+{\ln \delta \over12}\left(\chi[M_B^\alpha]
	+{1 \over 2\alpha}(1-\alpha)^2\right)~~~,
	\end{equation}
	\begin{equation}\label{sing-eul}
	\chi[M_B^\alpha]={1 \over 4\pi}\left(\int_{M_B^\alpha} Rd^2x
+2\int_{\Sigma_B}
k dy +4\pi(1-\alpha)\right)~~~,
	\end{equation}
where, as in (\ref{2.7a}), $\delta$ is the ultraviolet cut-off
parameter, $R$
is the regular curvature. The quantity $\chi[M_B^\alpha]$ is the
Euler
characteristic of $M_B^\alpha$ and it is identical to that of the
Gibbons-Hawking instanton\cite{FS1}: $\chi[M_B^\alpha]=\chi[M_B]=1$.
Thus, up
to the terms $(1-\alpha)^2$ the divergences on a regular instanton
and on a
singular one coincide (compare (\ref{2.7a}) and (\ref{divsing})) and
the
difference between them, being taken on-shell, does not affect the
entropy. As
earlier we assume that the renormalization has been  already done and
further
we use only renormalized quantities.

Let us calculate the off-shell effective action  $W_1^{CS}$ and
entropy
$S_1^{CS}$ by the conical singularity method.

\begin{figure}
\label{f4}
\let\picnaturalsize=N
\def\picsize{8cm}
\def\picfilename{fig4.eps}
\ifx\nopictures Y\else{\ifx\epsfloaded Y\else\input epsf \fi
\let\epsfloaded=Y
\centerline{\ifx\picnaturalsize N\epsfxsize \picsize\fi
\epsfbox{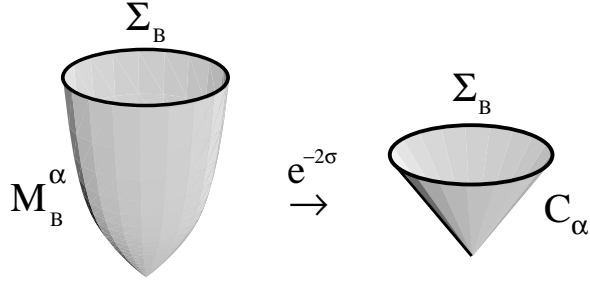}}}\fi
\caption[f4]{Conformal map of a singular instanton $M_\beta^\alpha$
onto  the
standard cone $C_{\alpha}$.}
\end{figure}
As earlier  $\beta$ is
the inverse temperature on $\Sigma_B$ and $\alpha={\beta_\infty
/\beta_H}$ is
the off-shell parameter.
We again use the conformal transformation (\ref{2.71}), but now it
maps a
singular instanton onto the standard cone $C_{\alpha}$
with the unit (in the units of $\mu$) length of the generator
	\begin{equation}\label{5.1}
	d\tilde{s}^2=\mu^2(x^2d\tilde{\tau}^2+dx^2)~~~,~~~0\leq x\leq
1~~~,~~~0\leq\tau\leq 2\pi\alpha~~~.
	\end{equation}
Eqs.(\ref{2.71}),
(\ref{2.73}) and (\ref{a10}) enable one to relate the effective
action
$W_1^{CS}$ to the action on
$C_{\alpha}$. Written as earlier in terms of variables
$(\beta,\alpha,y)$
this action takes the form
	\begin{equation}\label{5.2}
	W_1^{CS}(\beta,r_B,r_+)=
       \tilde{W}_1^{CS}(\beta,\alpha(\beta,r_B,r_+),
y(r_B,r_+))~~~.
	\end{equation}
	\begin{equation}\label{5.3}
	\tilde{W}_1^{CS}(\beta,\alpha,y)=
	-{\alpha \over 48}\left(2y+13y^2-15+4\ln{\beta\over
2\pi\mu\alpha}\right)
	\end{equation}
	\[
	-{1 \over 24\alpha}\left(\frac 1y -1 -\ln y +2\ln {\beta\over
2\pi\mu\alpha}\right)
	+ C(\alpha )~~~.
	\]
Here $C(\alpha)$ is the effective action for the unit cone which for
$\alpha=1$
coincides with the effective action on the unit disk $D^2$ denoted
earlier as
$C$: $C(\alpha=1)=C$. The function $C(\alpha)$ does not depend on
$\mu$ and
results in a numerical addition to the entropy. Its form is not
important for
our consideration.

For the on-shell limit $\alpha=1$ the cone singularity disappears, so
that one
has
	\begin{equation}\label{5.4}
	\tilde{W}_1^{CS}(\beta,\alpha=1,y)=\tilde{W}_1(y,\beta)~~~,
	\end{equation}
where $\tilde{W}_1(y,\beta)$ is the on-shell effective action given
by
Eq.(\ref{2.75a})

The entropy $S_1^{CS}$ is defined from
$\tilde{W}_1^{CS}(\beta,\alpha, y)$ by
the equation (\ref{entr2}). The calculation gives
	\begin{equation}\label{5.5}
	S_1^{CS}(\beta,\alpha,y)={1 \over 12\alpha}\left(\frac
1y-1-\ln y + 2\ln
{\beta 	\over 2\pi\mu\alpha}\right)+ C^{CS}(\alpha)~~~,
	\end{equation}
where
	\begin{equation}\label{5.6}
	C^{CS}(\alpha)=(\alpha {\partial \over \partial
\alpha}-1)C(\alpha)
	\end{equation}
is an irrelevant constant at $\alpha=1$.
Note that in the conical singularity approach both the renormalized
action
$W_1^{CS}$ and the entropy $S_1^{CS}$ are finite quantities.

\section{Blunt Cone Method}
\setcounter{equation}{0}
Consider as earlier the singular instanton $M_B^\alpha$ shown in
Fig.4  and a
set of regular manifolds that modify its geometry in the narrow
vicinity of the
sharp cone vertex (see Fig 5). The Riemann curvature for such
geometries is
regular everywhere and it differs from the Riemann curvature on a
singular
instanton only near the horizon.
We call this geometry the "blunt instanton"
and refer to this  off-shell extension\cite{a3,FS1} as to the {\em
blunt cone}
method. In  this approach we can avoid the problems connected with
the
formulation of the quantization and renormalization procedures on
manifolds
with infinite curvatures.
The regularization of the cone singularity is supposed to be removed
at the
very end of calculations.

      For simplicity of calculations we choose a special form of the
off-shell
extension  characterized by only two parameters: an off-shell
parameter
$\alpha=\beta_{\infty}/\beta_{H}$ and a new parameter $\eta$ which
describes
the width of the rounded tip of the blunt instanton.
We choose the metric on a blunt instanton  in the form
	\begin{eqnarray}\label{6.1}
	d s^2 &=& \left( {\beta \over 2 \pi} \right)^2
	\left( \rho^2 d\tau^2 + b^2 d \rho^2 \right) ; \hspace{1cm}
0 \leq \tau\leq
2	\pi,\hskip .7cm 0 \leq \rho\leq 1~~~;        \label{Blunt} \\
 	b &=& {1 \over (1-  \rho^2 + y \rho^2)^2 }  {\rho^2 +\alpha
\eta^2  \over
\alpha \rho^2 + \alpha \eta^2}~~~. \nonumber
	\end{eqnarray}
The boundary $\Sigma_B$ of the region under consideration is located
at
$\rho=1$, and its length  is $\beta$.  The parameter of the black
hole mass
enters, as earlier, through the dimensionless quantity $y=r_+/r_B$.
The
parameters that uniquely fix a {\em blunt instanton} are
$\beta,r_B,r_+$ and
$\eta$.
For $\alpha =  1$ the metric  is  identical to the metric of the
Gibbons-Hawking instanton.

\begin{figure}
\label{f5}
\let\picnaturalsize=N
\def\picsize{5.5cm}
\def\picfilename{fig5.eps}
\ifx\nopictures Y\else{\ifx\epsfloaded Y\else\input epsf \fi
\let\epsfloaded=Y
\centerline{\ifx\picnaturalsize N\epsfxsize \picsize\fi
\epsfbox{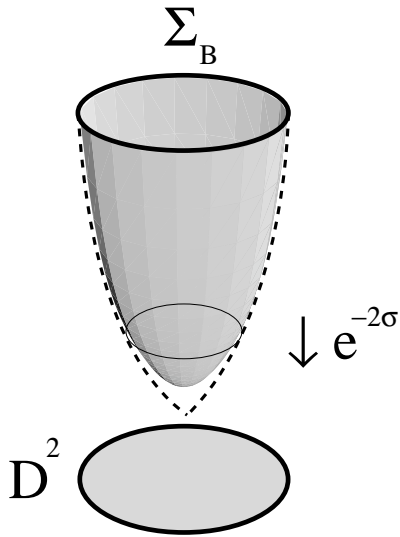}}}\fi
\caption[f5]{Blunt instanton and its conformal transformation onto a
unit disk
$D^2$}
\end{figure}

To calculate the renormalized one-loop  effective action on the blunt
instanton
we map the latter onto  a unit disk $D^2$.
Consider at first an arbitrary static Euclidean 2-D manifold with the
line
element $ds^2$ that is conformally related to the unit disk with the
element
$d\tilde{s}^2$
	\begin{eqnarray}
	ds^2 &=&\left( {\beta \over 2 \pi} \right)^2 \ [ a^2 d\tau^2
+ b^2 d\rho^2 ]
	= \exp (2\sigma) \ \mu^2 \ [ x^2 d\tilde{\tau}^2 + d x^2
]~~~,
\label{blumetric}
	\end{eqnarray}
where $0\leq \tau\leq 2\pi$,  $0\leq \tilde{\tau} \leq 2\pi$, $0\leq
\rho\leq
1$, and
$0\leq  x \leq 1$.
Then the metric coefficients $a,b$,  and the conformal factor
$\sigma$
	\begin{equation}\label{6.1a}
	\sigma (\rho) = \ln {a(\rho) \over a(1)} + \int_\rho^1 d \rho
{b \over a}
	+ \ln  \left( {\beta \over 2 \pi \mu} \right)
	\end{equation}
depend only on $\rho$.
The normalization of $\sigma$ is fixed by the requirements  $\sigma
(1) = \ln
\left( {\beta / 2 \pi \mu} \right)$ and $\tilde{\tau} = \tau$.

Integration of the conformal anomaly (see appendix A), when applied
to the
metric (\ref{blumetric}) gives
the  one-loop effective action
	\begin{eqnarray}
	W^{BC}_1  &=& - {1 \over 6} \ln  \left( {\beta \over 2 \pi
\mu} \right)
	- {1 \over 12}
	\int_0^1 d \rho { ( a' - b )^2 \over a b }
	- \left({a' \over 4 \ b}\right)_{\rho = 1}
	+ { 1\over 4}  + C~~~.
	\end{eqnarray}
Here $a'=da/d\rho$ and the constant $C$ is as earlier the effective
action
for the unit disk $D^2$.
To derive this formula the regularity condition $(a ' / b)\mid_{\rho
= 0} = 1$
of the metric at the horizon has been used. For the metric
(\ref{6.1}) of the
blunt instanton one has
	\begin{eqnarray}
	a &=& \rho ~~~, \hskip 2cm
	b = {1 \over (1-  \rho^2 + y \rho^2)^2 }  {\rho^2 +\alpha
\eta^2  \over
\alpha \rho^2 + \alpha \eta^2}~~~,  \\
	\sigma &=& \ln \rho + {1 \over 2} \int_{\rho^2}^1 dz
	{ z + \alpha\eta^2 \over z (\alpha z + \alpha \eta^2)(1- z +
y z)^2} +
	\ln  \left( {\beta \over 2 \pi \mu} \right)~~~,
\nonumber
	\end{eqnarray}
and the blunt cone effective action $W_1^{BC}$ reads
	\begin{equation}\label{6.1c}
	W_1^{BC} (\beta,r_B,r_+,\eta) =
	\tilde{W}_1^{BC}
(\beta,\alpha(\beta,r_B,r_+),y(r_B,r_+),\eta)~~~,
	\end{equation}
	\begin{eqnarray}
	\tilde{W}_1^{BC} (\beta,\alpha,y,\eta)&=&
	 - {1 \over 6} \ \ln \left[  {\beta \over 2\pi \mu} \right]
	- {(\alpha - 1) \over 24 \  \alpha}
	{1 \over (1+ \eta^2 - y \eta^2)^2}
	\ln \mid {\eta^2 \over 1+ \eta^2} \mid  \nonumber \\
	&+&  {\alpha - 1 \over 24} \ (1+ \alpha  \eta^2 - y \alpha
\eta^2)^2
	\ln \mid {\alpha \eta^2  \over 1+\alpha \eta^2 }\mid
	  \nonumber \\
	&+& {1 \over 24 } \ln \mid y  \mid \left\{ 1 - {\alpha -1
\over \alpha}
	{1 \over (1+\eta^2 - y \eta^2)^2} \right\}  \\
	&+& {1 \over 24} (1-y) \left\{ 2 \alpha -
	{1+\alpha  \eta^2 - y \alpha  \eta^2\over \alpha y
	(1+\eta^2 - y \eta^2) }\right\}   \nonumber \\
	&-&{ 1\over 48} \alpha (1-y)^2 \left\{ 1 - 2 (\alpha-1)
\eta^2 \right\}
	- {1 \over 4} {\alpha + \alpha \eta^2  \over 1 + \alpha
\eta^2 }
	y^2 + {1 \over 4}  + C ~~~.      \nonumber
	\end{eqnarray}
The parameter $\eta$ in the blunt-cone method plays the role similar
to the
cut-off parameter $\epsilon$ in the brick-wall method.
When the regularization parameter $\eta$ tends to zero $\eta
\rightarrow 0$,
the action becomes
	\begin{eqnarray}
	\tilde{W}_1^{BC} (\beta,\alpha,y,\eta) &=&
	- {1 \over 6} \ln {\beta \over 2\pi \mu}
	+ {1 \over 48 } \left[
	- { 2 \over \alpha y} + {2 \over \alpha} \ln y - 2 \alpha y -
13 \alpha y^2
	 \right.  \nonumber \\
	&+& \left. 2 (\alpha -1) \ln \alpha + {2 \over \alpha} + 3
\alpha + 12
	\right]  + C \\
	&+& {1 \over 24 \alpha}(\alpha - 1)^2 \ln \eta^2 +
O(\eta^2)~~~.
	\nonumber
	\end{eqnarray}
The metric (\ref{Blunt}) on-shell ($\alpha=1$)  becomes  the metric
of the
Gibbons-Hawking instanton and the corresponding on-shell effective
action reads
	\begin{eqnarray}
	\tilde{W}_1^{BC} (\beta,\alpha=1,y,\eta) &=&
	- {1 \over 6} \ln {\beta \over 2\pi\mu}
	+ {1 \over 48 }\left[ - {2 \over y}+2 \ln y-2y-13  y^2
        +17 \right] + C~~~.
	\end{eqnarray}
It is identical to the on-shell action $\tilde{W}_1(\beta,y)$ given
by
expression (\ref{2.75a}).
The  corresponding blunt-cone entropy remains finite in the limit
$\eta = 0$
and reads
	\begin{eqnarray}
	S_1^{BC} (\beta,1,y,0) &=& {1 \over 12 \ y}
	- {1 \over 12 } \ln  y
	+ {1 \over 6} \ \ln  {\beta \over 2\pi\mu} - {1 \over 2 }
	-  C ~~~.
	\end{eqnarray}
This result coincides (up to an unimportant constant) with the
entropy
$S_1^{CS}$ found by the conical singularity method.

\section{Method of the volume cut-off}
\setcounter{equation}{0}
Finally we discuss here one more method of the off-shell
definition of the black hole effective action $W_1$.
Note that $W_1$ can be represented as the volume integral over the
background
space of some Lagrange density ${\cal L}_1(x)$
\begin{equation}\label{7.1}
W_1=\int \sqrt{g}dx{\cal L}(x)~~~.
\end{equation}
The corresponding density ${\cal L}_1(x)$ can be written in terms of
the
diagonal elements of the heat kernel operator in the coordinate
representation
\begin{equation}\label{7.2}
{\cal L}_1(x)=-\frac 12 \int_{0}^{\infty}{ds \over s}\langle
x|e^{s\triangle}|x\rangle ,
\end{equation}
so that for the action itself one has the standard formula
\begin{equation}\label{7.3}
W_1=\frac 12 \ln \det(-\mu^2\triangle)=-\frac 12 \int_{0}^{\infty}
{ds \over
s}Tr~e^{s \mu^2\triangle}~~~.
\end{equation}

\begin{figure}
\label{f6}
\let\picnaturalsize=N
\def\picsize{10cm}
\def\picfilename{fig6.eps}
\ifx\nopictures Y\else{\ifx\epsfloaded Y\else\input epsf \fi
\let\epsfloaded=Y
\centerline{\ifx\picnaturalsize N\epsfxsize \picsize\fi
\epsfbox{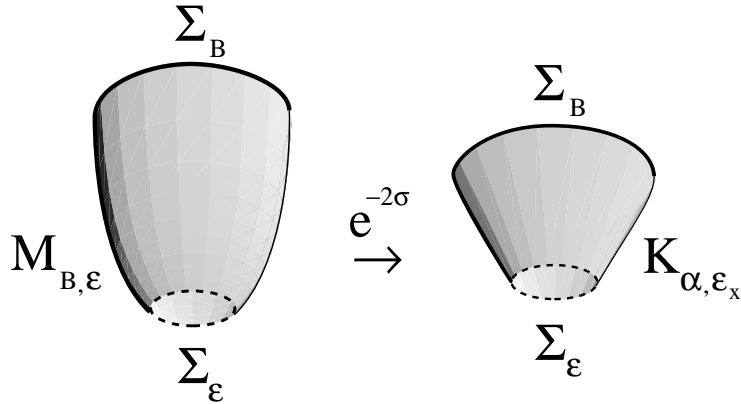}}}\fi
\caption[f6]{Volume cut-off}
\end{figure}

Consider now a singular instanton, and calculate  ${\cal L}_1(x)$ for
its
regular points $r>r_+$.  Denote by   $\Sigma_\epsilon$ a surface
located at a
small  proper distance $\epsilon$ from the horizon, and restrict the
integration   in Eq.(\ref{7.3}) by the region $M_{B,\epsilon}$
located outside
$\Sigma_\epsilon$, see Fig.6. As the result the action $W_1$  depends
on a new
parameter $\epsilon$. We call this off-shell
procedure the  {\em volume cut-off} method and denote the
corresponding
quantities with the superscript $VC$.

The volume (or spatial) cut-off method arises naturally in the
dynamical-interior approach to the black-hole entropy, proposed in
Ref.\cite{FrNo:93}. In this approach the internal degrees of freedom
of a black
hole are identified with the states of fields propagating in its
interior in
the close vicinity to the horizon. Because of the quantum
fluctuations of the
horizon, the separation of the modes into external (propagating
outside the
horizon) and internal (propagating inside the horizon) becomes
impossible for
modes located closer  to the horizon than the amplitude of  its
quantum
fluctuations. For this reason the summation of the modes which
contribute to
the statistical-mechanical entropy of a black hole in the approach
\cite{FrNo:93} is restricted only to the modes, which are located
outside the
fluctuation region of the horizon. This is equivalent to the spatial
cut-off in
the volume integral for the effective action described above.
The volume cut-off procedure has been also used in many other
papers\cite{DowkerCQG,BFZ,Barb,Emp,DeAlwis,CVZ,BCZ,BB}.
In works\cite{BFZ,Emp,DeAlwis,CVZ,BCZ,BB} the black hole metric has
been mapped
onto an optical (ultrastatic) metric. The horizon then maps to
infinity and the
proper volume
of the optical space becomes infinite.  In order to deal with this
divergence
it is
natural to restrict the volume integration by a finite region. This
approach
enables one to get a number of interesting results for the entropy
corrections
even for the massive fields in spaces with the dimension larger than
two\cite{CVZ,BCZ} and for conformal fields with non-zero
spins\cite{BB}.

Up to a certain  extent the volume-cut-off method resembles   the
brick-wall
approach.  However they are certainly different because the volume
cut-off
method does not require any special boundary conditions on
$\Sigma_\epsilon$.
It is also non-sensitive to  the behavior of the quantum field in the
region
lying  closer than $\Sigma_\epsilon$  to the horizon.

The calculation of the Lagrangian ${\cal L}_1$ on the off-shell black
hole
solution can again
be carried out with the help of the conformal transformation
to the conical space. Using (\ref{a10}) one can write
the following relation
\begin{equation}\label{7.4}
{\cal L}_1=e^{-2\sigma}{\cal L}_1(C_{\alpha})-
{1 \over 24\pi}\left(
R\sigma-(\nabla \sigma)^2
+(2K\sigma+3\sigma_{,\mu}n^{\mu})\delta(r,r_B)\right)
\end{equation}
between ${\cal L}_1$ and the Lagrangian ${\cal L}_1(C_{\alpha})$
on a unit cone $C_\alpha$ valid in the region outside the horizon.
Here
$\delta(r,r_B)$ is the invariant delta function which is included to
reproduce
the surface terms
on the external boundary in the action. The factor $\sigma$ is given
by
Eq.(\ref{2.73}). Note that the terms in Eq.(\ref{a10}) which are
determined by
the value of the conformal parameter $\sigma$ on the cone apex do not
contribute to $W^{VC}_1$ in Eq.(\ref{7.4}).

To find ${\cal L}_1(C_{\alpha})$ one can use the Sommerfeld
representation for
the heat kernel \\
$K_\alpha(x,x')=~<x|e^{s\triangle}|x'>$ of the Laplace operator
on the conical space (\ref{5.1})
	\begin{equation}\label{7.5}
	K_\alpha(x,x',\tilde{\tau}-\tilde{\tau}')=
K(x,x',\tilde{\tau}-\tilde{\tau}')+
{i\over 	4\pi\alpha}\int_{\Gamma}\cot\left(
	{w \over
2\alpha}\right)K(x,x',\tilde{\tau}-\tilde{\tau}'+w)dw
	\end{equation}
relating it to the heat kernel $K(x,x',\tilde{\tau}-\tilde{\tau}')$
on a unit
disk $D^2$.
Here the integration contour $\Gamma$ lies in the complex plane and
consists of
two curves, going from $\mp\pi-(\tilde{\tau}-\tilde{\tau}')\pm
i\infty$ to
$\mp\pi-(\tilde{\tau}-\tilde{\tau}')\pm i\infty$ and intersecting the
real axis
between the poles of the integrand $-2\pi\alpha,~0$ and $2\pi\alpha$.
A derivation and discussion of this formula can be found in
\cite{Dowker:77,Deser:88,DF:94,CKV}.
The Lagrange density on a cone can be easily calculated if one
substitutes
(\ref{7.5}) in (\ref{7.3}). The result has a simple form
	\begin{equation}\label{7.6}
	{\cal L}_1(C_\alpha)={\cal L}_1(D^2)-{1 \over 24\pi
x^2}\left({1 \over
\alpha^2}-1\right)~~~.
	\end{equation}
Here ${\cal L}_1(D^2)$ is the Lagrange density on the unit disk
$D^2$. In what
follows we  omit ${\cal L}_1(D^2)$ which results in an irrelevant
constant in
$W_1^{VC}$.
The second term, vanishing at $\alpha =1$, arises as the result of
integration
in Eq.(\ref{7.5}). In the calculations
the integral over $s$ is taken first and then the formula
	\begin{equation}\label{7.7}
	{i \over 8\pi\alpha}\int_{\Gamma}{\cot\left({w/ 2 \alpha}
\right)
	\over \sin ^2w/2}dw=\frac 16\left({1 \over \alpha^2}-1\right)
	\end{equation}
is used.

Let $W_{1}^{VC}[C_{\alpha}]$ be the effective action on a cone
$C_\alpha$
obtained by the integration of Eq.(\ref{7.6})
till the point $x=\epsilon_x$. As earlier $\epsilon_x$ is related
with the
invariant distance $\epsilon$ to the horizon by Eq. (\ref{dist}).
This
functional reads
	\begin{equation}\label{7.8}
	W_{1}^{VC}[C_{\alpha}]=
	{1 \over 12}\left(\alpha -{1 \over \alpha}\right)\ln{
 	\epsilon_x^{-1}}~~~.
	\end{equation}
Then, by using the Eqs. (\ref{7.6}) and (\ref{dist}), one can write
the
complete effective action in the volume cut-off method as
	\begin{equation}
	W_1^{VC}(\beta,r_B,r_+,\epsilon)=W_1^{VC}[C_{\alpha}]-
	{1 \over 24\pi}\left(\int_{M_{B,\epsilon}}
	(R\sigma-(\nabla \sigma)^2)
	+\int_{\Sigma_{B}}(2K\sigma+3\sigma_{,\mu}n^{\mu})\right).
	\label{7.9a}
	\end{equation}
So eventually we have
	\[
	W_1^{VC}(\beta,r_B,r_+,\epsilon)=
	\tilde{W}_1^{VC}(\beta,\alpha(\beta,r_B,r_+),y(r_B,r_+),
\epsilon)~~~,
	\]
	\begin{equation}\label{7.9}
	\tilde{W}_1^{VC}(\beta,\alpha,y,\epsilon)={1 \over
12}\left(\alpha-{1 \over
\alpha}\right)\left(\ln{\mu \over \epsilon}-\ln{2\pi\mu\alpha \over
\beta}
	- \frac 12 \ln y - \frac 12 +{1 \over 2y}\right)
	\end{equation}
	\[
	+{\alpha \over 48 \pi}\left(-\frac 2y +2\ln y
-2y-13y^2+17+8\ln{2\pi\mu\alpha
\over \beta}\right) + o(\epsilon)~~~.
	\]
When taken on-shell ($\alpha=1$) the divergence $\ln \epsilon$ of
this
functional disappears and $\tilde{W}_1^{VC}$ coincides with the
action
(\ref{2.75a}) on the regular space
	\begin{equation}\label{7.10}
	\tilde{W}_1^{VC}(\beta,\alpha=1,y,\epsilon)=
\tilde{W}_1(\beta,y)~~~.
	\end{equation}
The entropy $S_1^{VC}(\beta,\alpha,y,\epsilon)$ calculated from the
action
(\ref{7.9}) reads
	\begin{equation}\label{7.11}
	S_1^{VC}(\beta,\alpha,y,\epsilon)={1 \over
12\alpha}\left(2\ln{\mu \over
\epsilon}+2\ln{\beta \over 2\pi\alpha}
	-\ln y -1 +\frac 1y\right)~~~.
	\end{equation}
On-shell $S_1^{VC}$ differs from the conical-singularity entropy
$S_1^{CS}$
only by a singular term
depending on $\epsilon$
\begin{equation}\label{7.12}
S_1^{VC}(\beta,\alpha=1,y,\epsilon)=
S_1^{CS}(\beta,\alpha=1,y)+\frac
16\ln{\mu
\over \epsilon}~~~.
\end{equation}
The entropy $S_1^{VC}$ can be also written as
	\begin{equation}\label{7.13}
	S_1^{VC}(\beta,\alpha,y,\epsilon)={1 \over 6\alpha}\ln{
	\epsilon_x^{-1}}~~~.
	\end{equation}
So this quantity coincides with the entropy computed
from the action $W_1^{VC}(C_\alpha)$. The coincidence takes place
because the
anomaly, which differs $W_1^{VC}(\beta,\alpha,y,\epsilon)$ from
$W_1^{VC}(C_\alpha)$, is proportional to $\beta$ and does not
contribute into
$S_1^{VC}$.

Another observation is that $S_1^{VC}$ coincides with the thermal
entropy of
the quantum gas in the volume of the size $\ln \epsilon_x^{-1}$. The
volume
cut-off entropy does
not contain
the term $\ln\ln\epsilon^{-1} $ which is present in the brick wall
entropy
$S_1^{BW}$ since the boundary condition on the quantum field at
$\Sigma_\epsilon$ is not imposed, and the field can freely fluctuate
on this
boundary, see  Appendix C.

\section{Off-Shell versus On-Shell}
\setcounter{equation}{0}
\subsection{Off-Shell and On-Shell Effective Actions}
In this Section we discuss and compare the results of the off-shell
and
on-shell calculations of the thermodynamical characteristics of a
black hole.
We begin by discussing the obtained results for the effective action.
It is convenient to introduce the following notation
	\begin{eqnarray}
	U(\beta,\alpha,y)&=& - {1 \over 6} \ln \left[{ \beta \over 2
\pi \mu} \right]
	+ {1 \over 48}\left[ - {2 \over y} + 2 \ln y + 17 - 2 y - 13
y^2 \right]
\nonumber \\
	&+& {\alpha -1 \over 48 \alpha}\left(
	{2 \over y} - 2 \ln y - 2 + 15 \alpha - 2 \alpha y - 13
\alpha y^2
	 \right)  \label{9.1} \\
	&-& { (\alpha - 1)^2 \over 12 \alpha} \ln\left[{ \beta \over
2 \pi  \mu}
	\right]+\left(\alpha +{1 \over \alpha}\right)\ln\alpha
{}~~~.\nonumber
	\end{eqnarray}
Then the one-loop contributions to the effective action calculated by
different
off-shell methods can be presented in the following form
	\begin{equation}
	\tilde{W}_1^{CS}(\beta,\alpha,y)=U(\beta,\alpha,y) +
C(\alpha)~~~, \label{9.2}
	\end{equation}
	\begin{equation}
	\tilde{W}_1^{BW}(\beta,\alpha,y,\epsilon)=U(\beta,\alpha,y) +
	 {1 \over 12}\left( \alpha + {1 \over \alpha} \right)
	\ln \left( {\epsilon \over \mu} \right)
	-\frac 12\ln{\pi\alpha \over \ln{(\beta/
2\pi\alpha\epsilon)}} ~~~,\label{9.3}
	\end{equation}
	\begin{equation}
	\tilde{W}_1^{BC}(\beta,\alpha,y,\eta) =U(\beta,\alpha,y)
	 + { (\alpha - 1)^2 \over 12 \alpha} \ln\left[{ \eta \beta
\over 2 \pi \alpha
\mu}
	\right] + { \alpha -1 \over 24 } \ln \alpha - {\alpha -5\over
4} + C ~~~,
\label{9.4}
\end{equation}
\begin{equation}
	\tilde{W}_1^{VC}(\beta,\alpha,y,\epsilon)=U(\beta,\alpha,y)
	- {1\over12}\left(\alpha-{1\over\alpha}\right)
\ln{\epsilon\over\mu}
{}~~~.\label{9.5}
\end{equation}
Here we again use the notations $y=r_+/r_B$ and
$\alpha(\beta,r_B,r_+)=\beta/
\left(4\pi r_+\sqrt{1 - {r_+ / r_B}}\right)$. The constants $C$ and
$C(\alpha)$
which enter these relations are the effective actions $W_1=\frac 12
\ln\det(-\mu^2\triangle)$ on the unit disk $D^2$ and on the unit cone
$C_\alpha$ respectively.

In the same notations the on-shell one-loop effective action is
\begin{equation}\label{9.6}
\tilde{W}_1(\beta,y)=U(\beta,\alpha=1,y)+C~~~ .
\end{equation}
A simple comparison of Eqs.(\ref{9.3}) and (\ref{9.4}) with
Eq.(\ref{9.6})
shows that
\begin{equation}\label{9.7}
\tilde{W}_1^{CS}(\beta,\alpha=1,y)=
\tilde{W}_1^{BC}(\beta,\alpha=1,y,\eta)=
\tilde{W}_1^{VC}(\beta,\alpha=1,y,\epsilon)=\tilde{W}_1(\beta,y)~~~ .
\end{equation}
In other words the on-shell values of the one-loop effective actions
calculated
by conical singularity, blunt cone, and volume cut-off methods
coincide with
the on-shell one-loop effective action $\tilde{W}_1(\beta,y)$.
$\tilde{W}^{CS}_1$ is always finite, while $\tilde{W}_1^{BC}$ and
$\tilde{W}_1^{VC}$ are finite (i.e., do not contain either $\ln\eta$
or
$\ln\epsilon$ divergence) only on shell (for $\alpha=1$). The only
divergent
on-shell quantity is the brick wall effective action
$\tilde{W}_1^{BW}$.

The relation (\ref{9.3}) can be interpreted in the following way.
Let us remind that the effective action $W_1^{CS}$ has been
computed by the conformal map onto the cone $C_\alpha$, (see
Eq.(\ref{5.1})).
So $W_1^{CS}$ is defined up to
addition of the action $W_1[C_\alpha]=C(\alpha)$. Alternatively, one
could use
the map onto a cone $C_{\alpha,\epsilon}$ with the size
$\epsilon$.
The results of two computations can be compared
by using the difference between $W_1[C_\alpha]$ and
$W_1[C_{\alpha,\epsilon}]$.
This difference can be easily found because
both cones are related by the trivial rescaling:
\begin{equation}\label{22}
ds^2(C_{\alpha})=\left({\mu \over
\epsilon}\right)^{2}ds^2(C_{\alpha,\epsilon})~~~.
\end{equation}
Then Eq.(\ref{a10}) gives
\begin{equation}\label{24}
W_1[C_{ \alpha}]=W_1[C_{\alpha,\epsilon}]+{1 \over 12 }\left({1 \over
\alpha}+\alpha\right)\ln{\epsilon \over \mu}~~~.
\end{equation}
It enables one to represent the result (\ref{9.3}) as
\begin{equation}\label{9.13}
W_1^{BW}(\beta,\alpha,y,\epsilon)=
W_1^{CS}(\beta,\alpha,y)-W_1[C_{\alpha,\epsilon}]+
W_1^{CAS}(\beta,\alpha,\epsilon)~~~,
\end{equation}
where
\begin{equation}\label{Casimir}
W_1^{CAS}(\beta,\alpha,\epsilon)=-\frac 12\ln{\pi\alpha \over
\ln{(\beta/
2\pi\alpha\epsilon)}}
\end{equation}
is the contribution due to the Casimir effect.
The detail discussion of this term and its relation to the brick-wall
boundary
conditions is given in the Appendix C.

\subsection{Why the on-shell and off-shell one-loop contributions
to the entropy are different}

The equality (\ref{9.7}) of all (except brick wall) off-shell
effective
actions and the on-shell effective action does not guarantee that the
same is
true for the corresponding values of entropy. Moreover, as we shall
see all the
off-shell calculations give the results for the entropy which differ
from the
on-shell result. Before giving the concrete relations between these
quantities
let us discuss why it happens.

Our starting point in the off-shell calculations is the one-loop
action
$W^\bullet_1$ which is the function of the parameters $\beta$, $r_B$,
and
$r_+$.
In the 'brick wall' and volume cut-off approaches it also depends on
the
additional parameter $\epsilon$,  and on $\epsilon$ and $\eta$ in the
blunt
cone method. The dependence on these additional parameters is not
important at
the moment, so we will not indicate it explicitly.
The quantities $\beta$ and $r_B$ are external parameters fixing the
problem and
 $r_+$ is determined on-shell in terms of them  by the condition
	\begin{equation}\label{8a.1}
	\alpha(\beta,r_B,r_+)={\beta \over 4\pi r_+\sqrt{1 - {r_+ /
r_B}}}=1
	\end{equation}
Consider first cone-singularity, blunt cone, and volume cut-off
methods for
which the  effective actions, when taken on shell (\ref{8a.1}),
coincide with
the thermodynamical action $W_1(\beta,r_B)$ given by Eqs.(\ref{2.75})
and
(\ref{2.75a})
	\begin{equation}\label{8.2}
	\left. W_1^{~\bullet}(\beta,r_B,
r_+)\right|_{\alpha=1}=W_1(\beta,r_B)~~~ .
	\end{equation}
Here the symbol $\bullet$ replaces CS, BC and VC notations. The
thermodynamical
entropy $S^{TD}_1$ is defined by Eq.(\ref{2.75c})
	\begin{equation}\label{8.3}
	S^{TD}_1=\beta\left. {\partial {W_1(\beta,r_B)}
	\over \partial \beta}\right| _{r_B} -W_1(\beta,r_B)~~~ ,
	\end{equation}
while the off-shell entropy $S^{\bullet}_1$ is defined by
Eq.(\ref{entr1})
	\begin{equation}\label{8.4}
	S^{~\bullet}_1= \beta\left. {\partial
{W_1^{~\bullet}(\beta,r_B,r_+)}\over
\partial 	\beta}\right| _{r_B,r_+}
-W_1^{~\bullet}(\beta,r_B,r_+)~~~ .
	\end{equation}
Note that in the calculation of $S^{~\bullet}_1$ the parameter $r_+$
is assumed
to be fixed. This results in the difference $\Delta S^{~\bullet}$
between two
entropies
	\begin{equation}\label{8.6}
	\Delta S^{~\bullet}=
S_1^{TD}-S_1^{~\bullet}=\beta\left({\partial \over
\partial 	\beta}W_1(\beta,r_B)-\left.{\partial \over \partial
\beta}W_1^{~\bullet}(\beta,r_B,r_+)\right)\right|_{\alpha=1}~~~.
	\end{equation}
Together with the Eq.(\ref{8.2}) it gives
	\begin{equation}\label{8a.6}
	\Delta S^{~\bullet}=\beta\left( {\partial r_+ \over \partial
\beta} \left.
{\partial W_1^{~\bullet}\over \partial
	r_+}\right|_{\beta,r_B}\right)_{\alpha=1}~~~
	\end{equation}
which, obviously, is non zero quantity. This shows why in the general
case the
one-loop contribution to the black hole entropy found by an off-shell
procedure
 differs from the contribution inferred in the thermodynamical
computation,
based on the on-shell action.

\subsection{Relations between off-shell and on-shell entropies}

We obtain now explicit formulas relating different off-shell
entropies. As
earlier we assume that after the calculations of the entropy the
limit
$\alpha=1$ is taken. The calculated entropies are always understood
as the
function of the parameters $\beta,r_B$ characterizing the system. For
simplicity we omit these arguments. Note also, that  the effective
actions
contain an arbitrary constants, which we denoted as $C$ and
$C(\alpha)$. It is
evident that similar constants enter also the expressions for the
entropies. We
indicated these constants explicitly earlier in the expressions for
the
entropies. They may be important for the discussion of the questions
connected
with the third law of black-hole thermodynamics. But they are not
important for
us now. For this reason in order to simplify the expressions we
simply omit
them from now on. We also omit the terms which vanish when the
additional
parameters (such as $\epsilon$ and $\eta$) take their limiting value
($\epsilon=0$ and $\eta=0$).

It is convenient to begin with the entropy $S_1^{CS}$ calculated by
the conical
singularity method. It is obtained from the effective action
$W_1^{CS}$ given
by (\ref{9.2}) with $C(\alpha=1)=0$, or what is equivalent from $U$,
given by
Eq.({\ref{9.1})
\begin{equation}\label{9.21}
S_1^{CS}={1 \over 12}\left(\frac 1y-1-\ln y + 2\ln {\beta \over
2\pi\mu}\right)~~~.
\end{equation}
Let us denote
\begin{equation}\label{9.22}
S_1^T(\epsilon)=\frac 16\ln{\mu \over \epsilon} ,
\hspace{1cm}S_1^{CAS}(\epsilon)=\frac 12\ln{\pi \over \ln{\beta \over
2\pi\epsilon}} .
\end{equation}
Then the results of the previous sections can be summarized as
follows
\begin{equation}\label{9.24}
S_1^{BW}=S_1^{CS}+S_1^T+S_1^{CAS}~~~,
\end{equation}
\begin{equation}\label{9.25}
S_1^{VC}=S_1^{CS}+S_1^T~~~,
\end{equation}
\begin{equation}\label{9.26}
S_1^{BC}=S_1^{CS}~~~.
\end{equation}
Thus, the blunt cone and conical singularity methods give the same
finite
result for the entropy, while the brick wall and volume cut-off
methods give
expressions  containing ($\ln \epsilon$)-divergence. The difference
$S_1^{CAS}$
 between $S_1^{BW}$ and $S_1^{VC}$  occurs because the different
boundary
conditions in these methods are imposed. All the above off-shell
expressions
for the entropy differ from the one-loop contribution $S_1^{TD}$ to
the
thermodynamical entropy given by Eq.(\ref{2.76}). The latter can be
presented
as
\begin{equation}\label{9.18}
S_1^{TD}=S_1^{CS}+\Delta S ,
\end{equation}
where
\begin{equation}\label{9.19}
\Delta S\equiv \beta\left( {\partial r_+ \over \partial \beta}\left.
{\partial
W_1^{CS}\over \partial
	r_+}\right|_{\beta,r_B}\right)_{\alpha=1}= {1\over
48(2-3y)}(-14+26y-28y^2+13y^3)+{1\over 24}\ln y .
\end{equation}

The relation (\ref{9.24})   can be rewritten in a different form
which is more
convenient for interpretation. Note that according to
Eqs.(\ref{4.16}),(\ref{4.17}) and (\ref{4.21a})
\begin{equation}\label{9.30}
S_1^{BW}=- \mbox{Tr}\left(\hat{\rho}^{H}_{\epsilon}(\beta)\ln
\hat{\rho}^{H}_{\epsilon}(\beta)\right) .
\end{equation}
On the other hand $S_1^T+S_1^{CAS}$ can be identically rewritten as
\begin{equation}\label{9.31}
S_1^T+S_1^{CAS}=S_\epsilon^R(2\pi\mu)=
-\mbox{Tr}\left(\hat{\rho}_\epsilon^R(2\pi\mu)
\ln \hat{\rho}_\epsilon^R(2\pi\mu)\right)~~~.
\end{equation}
That is this expression coincides with the entropy of a massless
thermal
radiation in the Rindler space  between two mirrors located at the
proper
distances $\epsilon$ and $\mu$ from the horizon.
The temperature of the radiation measured at the distance $\mu$ from
the
horizon is $1/(2\pi\mu)$.
Thus we have
\begin{equation}\label{9.32}
S_1^{CS}=- \left[\mbox{Tr}\left(\hat{\rho}^{H}_{\epsilon}(\beta)\ln
\hat{\rho}^{H}_{\epsilon}(\beta)\right)-
\mbox{Tr}\left(\hat{\rho}^R_{\epsilon}(2\pi\mu)
\ln \hat{\rho}^R_{\epsilon}(2\pi\mu)\right)\right]~~~.
\end{equation}
It is easy to verify that the same relation is valid also if the
inner
mirror-like boundary  (at $\epsilon$)  is absent provided the
quantities in the
right-hand side are defined by using the volume cut-off method. For
both brick
wall and volume cut-off methods
each of the terms in the right-hand side of Eq.(\ref{9.32}) is
divergent as
$\epsilon\rightarrow 0$, while the difference remains finite in this
limit.
If we formally define
the density matrices $\hat{\rho}^{H}(\beta)$ and
$\hat{\rho}^R(2\pi\mu)$ on the
black-hole and Rindler backgrounds  as the limits
\begin{equation}\label{9.6d}
\hat{\rho}^{H}(\beta)=\lim_{\epsilon\rightarrow
0}\hat{\rho}_\epsilon^{H}(\beta)~~~,
{}~~~\hat{\rho}^R(2\pi\mu)=\lim_{\epsilon\rightarrow 0}
\hat{\rho}_\epsilon^R(2\pi\mu)~~~
\end{equation}
then for both volume cut-off and brick wall methods, we have
\begin{equation}\label{9.6c}
S_1^{CS}(\beta,\alpha=1,y)=-
\left[\mbox{Tr}\left(\hat{\rho}^{H}(\beta)\ln
\hat{\rho}^{H}(\beta)\right)-\mbox{Tr}\left(\hat{\rho}^R(2\pi\mu)\ln
\hat{\rho}^R(2\pi\mu)\right)\right]~~~.
\end{equation}
Using Eq.(\ref{9.18}) we finally get
\begin{equation}\label{9.38}
S_1^{TD}=-\left[\mbox{Tr}\left(\hat{\rho}^{H}(\beta)\ln
\hat{\rho}^{H}(\beta)\right)-\mbox{Tr}\left(\hat{\rho}^R(2\pi\mu)\ln
\hat{\rho}^R(2\pi\mu)\right)\right]+\Delta S .
\end{equation}
This relation indicates that the one loop correction to the
thermodynamical
entropy can be obtained from the statistical-mechanical black hole
entropy by
the following procedure.
First one needs to subtract the Rindler entropy which removes the
divergence,
and then add a finite correction $\triangle S$. In the next section
we show
that the second term  $\Delta S$ coincides with the change of the
classical
Bekenstein-Hawking entropy due to
the quantum deformation of the background geometry.

It is worth mentioning that a similar subtraction procedure naturally
arises in
the membrane paradigm\cite{ThPrMa:86}. Namely, in order to obtain the
correct
expression for the flux of the entropy
onto a black hole, Thorne and Zurek\cite{ZuTh:85,ThPrMa:86}
proposed to subtract from the entropy, calculated by a
statistical-mechanical
method, the entropy of a thermal atmosphere of the black hole. The
later
entropy close to the horizon coincides with $S^{SM}_{Rindler}$.
Eq.(\ref{9.38})
can be used to prove this conjecture. However, it should be stressed
that
Thorne and Zurek did not consider quantum corrections to the entropy
discussed
in the present paper. Eq.(\ref{9.38}) not only explains how the
volume
infinities in $S^{SM}$ are separated, but also gives an exact
dependence of the
quantum corrections
to the entropy on physical characteristics.

\subsection{Entropy and backreaction effects}

The thermodynamical entropy of a black hole with quantum one-loop
corrections
is
	\begin{equation}\label{9.40a}
		S^{TD}=S^{BH}(r_+)+S^{TD}_1~~~,
	\end{equation}
where $S^{BH}(r_+)=\pi r^2_+$ is the Bekenstein-Hawking entropy. As
the result
of quantum effects a 'real' solution $(\bar{\gamma},\bar{r})$
including quantum
corrections is different from the classical Schwarzschild solution
$(\gamma,r)$
\cite{Anderson}. In particular the value $\bar{r}_+$ of the dilaton
field at
the horizon of $\bar{\gamma}$ differs from its classical value $r_+$.
We
demonstrate now that Eq.(\ref{9.40a}) can be identically rewritten as
	\begin{equation}\label{9.41a}
		S^{TD}=\pi\bar{r}_{+}^2+S^{CS}_1~~~.
	\end{equation}

A first step in the proof is to obtain an  equation which determines
$\bar{r}_+$. For given boundary conditions $(\beta,r_B)$ the extremum
of the
Euclidean effective action $W$ defines a regular quantum solution.
This
solution can be obtained by solving the field equations
$\delta W/\delta \bar{\gamma}=\delta W/\delta \bar{r}=0$
and fixing an arbitrary constant which enters the solution by the
regularity condition on the horizon. This determines $\bar{r}_+$ as a
function
of $(\beta,r_B)$:
$\bar{r}_+=\bar{r}_+(\beta,r_B)$.
For any other choice of the constant the solution has a cone-like
singularity. We call such a singular solution a
{\em quantum singular instanton}. It obeys local field equations but
does not
provide a global extremum for $W$.
The quantum singular instanton is specified by $(\beta,r_B)$ and an
arbitrary
parameter $\bar{r}_+$.
We write the solution as
$\left(\bar{\gamma}(\bar{r}_+),\bar{r}(\bar{r}_+)\right)$.
The effective action $W(\beta,r_B,\bar{r}_+)$ calculated on the
quantum
singular instanton is
\[
W(\beta,r_B,\bar{r}_+)\equiv
W[\beta,r_B,\bar{\gamma}(\bar{r}_+),\bar{r}(\bar{r}_+)]=
\]
\begin{equation}\label{Q4a}
\hspace{2cm}I[\beta,r_B,\bar{\gamma}(\bar{r}_+),\bar{r}(\bar{r}_+)]+
W_1^{CS}[\beta,r_B,\bar{\gamma}(\bar{r}_+),\bar{r}(\bar{r}_+)]~~~.
\end{equation}
The condition of the global extremality of $W$
	\begin{equation}\label{9.41}
        {\partial {W(\beta,r_B,r_+)}\over \partial r_+}=0
	\end{equation}
determines the horizon radius $\bar{r}_+=\bar{r}_+(\beta,r_B)$ for
the regular
quantum instanton.

In the calculations we keep only terms up to the first order in
$\hbar$. For
this reason we can replace
$W_1^{CS}[\beta,r_B,\bar{\gamma}(\bar{r}_+),\bar{r}(\bar{r}_+)]$ in
the
right-hand side of Eq.(\ref{Q4a}) by its value calculated on the
classical
singular instanton
$W_1^{CS}[\beta,r_B,\gamma(\bar{r}_+),r(\bar{r}_+)]$.
What is much less trivial, we can also replace
($\bar{\gamma}(\bar{r}_+),\bar{r}(\bar{r}_+)$) in the classical
action $I$ in
(\ref{Q4a})
by the  solution  ($\gamma(\bar{r}_+),r(\bar{r}_+)$) for a classical
singular
instanton provided the value of the dilaton field $\bar{r}_+$ on the
horizon is
preserved the same.
To show this, consider the general variation of the classical action
$I$ given
by Eq. (\ref{2.3}). For   fixed $r_B$ and $\beta$ we have
\[
I[\beta,r_B,\bar{\gamma},\bar{r}]=I[\beta,r_B,\gamma,r]+\int
\left[\left.{\delta I \over
\delta\gamma_{ab}}\right|_{\gamma_{ab}}(\bar{\gamma}_{ab}-
\gamma_{ab})+{\delta
I \over \delta r}\delta r\right]
\]
\begin{equation}\label{Q4b}
+r_{,\mu}n^{\mu}|_{r=r_+}\delta r_+
-2\pi(1-\alpha)r_+\delta r_+ +O(\hbar ^2)~~~.
\end{equation}
We assume that the value of the dilaton field on the cone singularity
is $r_+$,
and denote by $2\pi(1-\alpha)$  the corresponding deficit angle which
is
defined by $(\gamma,r)$ at $r_+~$ \cite{fn7}.
The relation (\ref{Q4b}) shows that when the value $r_+$ for $\gamma$
and
$\bar{\gamma}$ is the same, and $(\gamma,r)$ is a solution of
classical
equations ( $\delta I / \delta\gamma_{ab}=0$, $\delta I / \delta
r=0$) the
value of the classical action calculated on ($\bar{\gamma},\bar{r}$)
differs
from the classical value $I[\beta,r_B,\gamma,r]$ only by terms of the
order
$O(\hbar^2)$.  That is why we can replace
$I[\beta,r_B,\bar{\gamma}(\bar{r}_+),\bar{r}(\bar{r}_+)]$ in
Eq.(\ref{Q4a}) by
$I(\beta,r_B,r_+)$, the value of $I$ calculated on the classical
singular
instanton. The latter can be easily found
	\begin{equation}\label{9.40}
		I(\beta,r_B,r_+)=\beta E(r_B,r_+)-\pi
r_+^2~,~~~E(r_B,r_+)\equiv
r_B\left(1-(1-r_+/r_B)^{1/2}\right),
	\end{equation}
where $E$ is a quasilocal energy \cite{York:86,BrownYork:93}.

The equation (\ref{9.41}) which defines the 'position' $\bar{r}_+$ of
the
quantum horizon
can be written as
	\begin{equation}\label{9.43}
		{\partial W_1^{CS}(\beta,r_B,r_+)\over \partial
r_+}=-2\pi\bar{r}_+
(\bar{\alpha}-1)~~~.
	\end{equation}
Here $\alpha=\alpha(\beta,r_B,r_+)=\beta[4\pi
r_+\sqrt{1-r_+/r_B}]^{-1}$, and
$\bar{\alpha}$ is the value of the classical off-shell parameter
$\alpha$
calculated for $r_+=\bar{r}_+$. For the classical regular instanton
$\alpha=1$.
It means that up to the second order in $\hbar$ we can write
	\begin{equation}\label{9.44}
		2\pi\bar{r}_+ (\bar{\alpha}-1)=2\pi
r_+\left({\partial{\alpha}\over{\partial{r_+}}}\right)_{\alpha=1}
\Delta r_+.
	\end{equation}
Here $\Delta r_+=\bar{r}_+ -r_+$ is the change of the 'position' of
the black
hole horizon because of the quantum corrections. Using the explicit
expression
for $\alpha$ it is easy to show that
	\begin{equation}\label{9.45}
\left({\partial{\alpha}\over{\partial{r_+}}}
\right)_{\alpha=1}=-\left[
\beta
{\partial {r_+}\over 		\partial
\beta}\right]^{-1}_{\alpha=1} .
	\end{equation}
The latter relation allows one to write
	\begin{equation}\label{9.46}
		2\pi r_+\Delta r_+=\beta\left[ {\partial {r_+}\over
\partial \beta}{\partial
{W_1^{CS}}\over \partial r_+}\right]_{\alpha=1},
	\end{equation}
and hence using Eq.(\ref{9.19}) one gets
	\begin{equation}\label{9.47}
		\Delta S =2\pi r_+\Delta r_+~~~.
	\end{equation}
Therefore, up to the terms $O(\hbar^2)$ the quantity $\Delta S$ can
be
represented as the difference $\Delta
S=S^{BH}(\bar{r}_+)-S^{BH}(r_+)$. On the
other hand, taking into account (\ref {9.18}), we can write the
thermodynamical
entropy given by Eq.(\ref{9.40a}) as the sum
$S^{TD}=S^{BH}(r_+)+\triangle
S+S_1^{CS}$. These equalities
prove the desired relation (\ref{9.41a}).

\section{Summary and Conclusions}
\setcounter{equation}{0}
Discuss now some lessons we have learned by comparing on-shell
results with the
results of the different off-shell methods in black hole
thermodynamics. First
of all direct calculations demonstrate that the thermodynamical
entropy of a
black hole $S^{TD}$ determined by the response of the free energy to
the change
of the temperature, and the statistical-mechanical entropy $S^{SM}$,
defined as
$S^{SM}=-\mbox{Tr}(\hat{\rho}^H\ln\hat{\rho}^H)$ for density matrix
$\hat{\rho}^H$ of black-hole internal degrees of freedom, are
different. The
thermodynamical entropy, besides the tree-level Bekenstein-Hawking
part
$S^{BH}=A/4$ contains also finite quantum one-loop correction
$S^{TD}_1$. The
latter can be obtained from the on-shell effective action. The
statistical-mechanical entropy $S^{SM}$ is defined as a one-loop
quantity, and
it requires an off-shell procedure for its calculation. $S^{SM}$ can
be
identified with the volume-cut-off entropy $S^{VC}_1$. Then it
contains the
divergence $(\ln \epsilon )$ where $\epsilon$ is a proper-distance
cut-off of
the volume integration, required to make this quantity finite. This
leading
logarithmical part of $S^{SM}$ also presents in the brick-wall
model, but
generally due to the Casimir effect, $S^{BW}_1$ has the additional
divergence
$(\ln |\ln \epsilon|)$.

The physical reason why $S^{TD}$ and $S^{SM}$ are different is
connected with a
special property of a black hole as a thermodynamical
system\cite{Frol:95}.
Namely, the internal degrees of freedom of a black hole are defined
as
excitations propagating on the back-ground geometry. This geometry is
uniquely
determined by the mass parameter, which in the state of thermal
equilibrium is
a function of the external temperature. For this reason,
to find $S_1^{TD}$ one must change the temperature. This results in
the change
of Hamiltonian, describing these internal excitations. On the other
hand, in
the calculations of $S^{SM}$ the black hole mass and the Hamiltonian
are to be
fixed.

We proved that the thermodynamical entropy of a black hole can be
presented in
the form
\begin{equation}\label{10.1}
S^{TD}=S^{BH}(\bar{r}_+)+\left[ S^{SM}-S^{SM}_{Rindler}\right] .
\end{equation}
$S^{BH}(\bar{r}_+)=\pi \bar{r}_+^2$ is the Bekenstein-Hawking
entropy, and
$\bar{r}_+$ is the 'radius' of the horizon of a 'quantum' black hole.
The term
in the square brackets is the difference between the
statistical-mechanical
entropies calculated for a black hole
$\left[S^{SM}=-\mbox{Tr}\left(\hat{\rho}^{H}(\beta)\ln
\hat{\rho}^{H}(\beta)\right)\right]$ and for a Rindler space
$\left[S^{SM}_{Rindler}=-\mbox{Tr}\left(\hat{\rho}^R(2\pi\mu)\ln
\hat{\rho}^R(2\pi\mu)\right)\right]$.
This {\em subtraction procedure} automatically removes all the
divergences from
$S^{SM}$ and results in an invariant regularization-independent
quantity.

We proved the relation (\ref{10.1}) by explicit calculations in 2-D
case, but
it seems to be of the general nature and it (or its generalization)
must be
valid in the 4-D case. The reason is that the on-shell renormalized
quantity
$S^{TD}$ is always finite, so that the subtraction terms  in
Eq.(\ref{10.1})
will always be of the form, required for the complete cancellation of
the
volume divergences of $S^{SM}~$ \cite{Frol:95}. One of the possible
ways to
derive in four dimensions the relation analogous to Eq.(\ref{10.1})
is to use
an optical metric, where the required subtraction terms can be
calculated by
using high-temperature expansion. For this reason, the coefficients,
which
enter the subtraction terms with different order of singularity in
$\epsilon$
must be connected with the Schwinger-DeWitt coefficients.

A remarkable property of the conical singularity method is that (at
least in
2-D case) it gives the finite result immediately
\begin{equation}\label{10.2}
S^{CS}_1=S^{SM}-S^{SM}_{Rindler} .
\end{equation}
The mathematical reason why $S^{CS}_1$ is finite while $S^{VC}_1$
contains
volume ($\ln \epsilon$) divergence is connected with the difference
of the
topologies of the
manifolds used to calculate the corresponding effective actions. For
$S^{VC}_1$
the standard manifold has the topology of a cylinder (or a ring),
while in for
$S^{CS}_1$ the topology is of $D^2$, i.e. the same as the topology of
the
Gibbons-Hawking instanton. The mathematical operation when one cuts a
small
disk of the radius $\epsilon$, from the standard unit disk $D^2$ to
transform
it into a ring, can be interpreted as the subtraction of an
entanglement
entropy\cite{Sorkin,Srednicki,KabStr,CalWil94,HLW:94}
$~S^{SM}_{Rindler}
=-{\mbox{Tr}}(\hat{\rho}^R\ln\hat{\rho}^R)$.

We stress once again that in our approach all the renormalizations
are to be
done from the very beginning so that only observable finite coupling
constants
enter the results. We demonstrated that some of the off-shell methods
require
an additional  cut-off parameter which we denoted by $\epsilon$. This
cut-off
parameter is completely independent from the ultraviolet cut-off
$\delta$, see
Eqs.(\ref{2.7a}) and (\ref{divsing}). Moreover the parameter
$\epsilon$ enters
only some intermediate quantities and never appears in the final
observable
results.
We demonstrated explicitly that quantum corrections to the physically
observable quantities can be always obtained by working only with
on-shell
quantities. As the result, for a black hole of a mass much greater
than the
Planckian mass the quantum corrections to observables are small and
independent
of the physics at Planckian scales. This differs on-shell quantities
from the
off-shell ones, such as $S^{SM}$.

There remains one more general question to be clarified.  All the
observables
characterizing a black hole in a thermal equilibrium, or its slow
transition
from one equilibrium state to another can be found by using only
on-shell
quantities. Why at all does one need to use off-shell methods in the
black hole
thermodynamics? We have already seen that one of the reasons is the
desire to
establish a relation between statistical-mechanical and
thermodynamical
entropies. In this sense, the off-shell methods can be considered as
a useful
tool for calculation  and  interpretation
of the on-shell quantities. But we believe that beside this trivial
reason
there may exist another more deep one.  The off-shell approaches may
also be
relevant for description of non-equilibrium processes in a system
including a
black hole. In this case quantum and thermal fluctuations of a
thermodynamical
system can be described by introducing stochastic noise\cite{Hu},
which
effectively takes a system off-shell. For this reason one may guess
that such
processes, for instance, as transition to a thermal equilibrium of a
black hole
initially exited by high energy explosion near its horizon may
require for
their consideration some of the above mentioned off-shell
characteristics.

\vspace{12pt}
{\bf Acknowledgements}:\ \ This work was supported  by the Natural
Sciences and
Engineering
Research Council of Canada.

\newpage
\appendix
\section{Conformal transformations of the effective action in two
dimensions}
\setcounter{equation}{0}
For completeness  we derive in this Appendix the conformal
transformations for the effective action
\begin{equation}\label{a1}
W_1[\gamma]=\frac 12 \ln \det[-\triangle]=-\frac 12
\int_{0}^{\infty}{ds \over
s} Tr(e^{s\triangle})
\end{equation}
defined on a 2-D Euclidean manifold ${\cal M}_{\alpha}$ with the
boundary $\partial {\cal M}_{\alpha}$ and a point
$x_s$ where ${\cal M}_{\alpha}$ has the conical singularity with the
deficit
angle $2\pi(1-\alpha)$. We will follow the method developed in
\cite{Dowker90}
and use
for this aim the dimensional regularization.
Consider the effective action $W_1$ for the conformally invariant
operator
$D=\triangle -(d-2)(4(d-1))^{-1} R$ in a $d$-dimensional space. The
divergent
part $W_1^{div}$ of $W_1$ can be found from the asymptotic heat
kernel
expansion
\begin{equation}
Tr(e^{sD})={1 \over (4\pi s)^{d/2}}\sum_{n=0,1/2,..}^{\infty}
a_n^{(d)}s^n~~~.
\label{a3}
\end{equation}
In 2-dimensional case for the dimensional regularization
\begin{equation}
W_1^{div}={1 \over d-2} {a_1^{(d)} \over 4\pi}~~~,
\label{a4}
\end{equation}
where  for an arbitrary $\alpha~$ \cite{DF:94(2),Dowker:9495}
\begin{equation}
a_1^{(d)}= \left(\frac 16-{d-2 \over 4(d-1)}\right) \int_{{\cal M
_{\alpha}}}
R+ {\pi \over 3}\left( {1 \over \alpha} -\alpha\right)\int_{\Sigma}
+\frac 13\int_{\partial {\cal M _{\alpha}}} k~~~.
\label{a5}
\end{equation}
In Eq.(\ref{a5}) the singular point $x_s$ is replaced by a singular
surface
$\Sigma$ of the dimension $d-2$ and
the integral of the scalar curvature $R$ is taken over the regular
part of
${\cal M} _{\alpha}$. $k$ is the second fundamental
form of the spatial boundary $\partial {\cal M}_{\alpha}$ defined in
terms of
its normal as $k=\nabla ^{\mu}n_{\mu}$.

The renormalized action is defined as the difference of the
non-renormalized
(bare) action $W_1^{bare}$ and its divergent part $W_1^{div}$
\begin{equation}
W_1=W_1^{bare}-W_1^{div}~~~.
\label{a6}
\end{equation}
Under conformal transformation $\tilde{\gamma}_{\mu\nu}=
e^{-2\sigma}\gamma_{\mu\nu}$ of the metric on ${\cal M}_\alpha$ the
renormalized action changes as\cite{Dowker90}
\begin{equation}
W_1(\tilde{\gamma})-W_1(\gamma)={1 \over 4\pi} \lim_{d\rightarrow 2}
{1 \over 2-d} ~\left( a_1^{(d)}(\tilde{\gamma})
-a_1^{(d)}(\gamma)\right)~~~.
\label{a7}
\end{equation}
Further we will consider only those transformations which do not
"squash" the
conical singularity. Then,
by making use of the following relations
\begin{equation}
\tilde{R}=e^{2\sigma}\left[R+(d-1)(2\triangle \sigma
+(2-d)\sigma_{,\alpha}\sigma^{,\alpha})\right]~~~,
\label{a8}
\end{equation}
\begin{equation}
\tilde{k}=e^{\sigma}(k-(d-1)\sigma_{,\mu}n^{\mu})
\label{a9}
\end{equation}
one gets from (\ref{a7})
\[
W(\tilde{\gamma})-W(\gamma)={1 \over 24\pi}\left[\int_{{\cal M_
{\alpha}}}
\left(R\sigma-(\nabla \sigma)^2\right) +\int_{\partial {\cal M _
{\alpha}}}(2k\sigma+3\sigma_{,\mu}n^{\mu})\right]
\]
\begin{equation}
+{1 \over 12}\left({1 \over \alpha} -\alpha \right)\sigma(x_s)~~~.
\label{a10}
\end{equation}
This is the desired conformal transformation of the effective action
where
$\sigma(x_s)$ is the value of the conformal factor in the point of
conical
singularity. If the manifold has a number of conical singularities in
points
$x_s$ with different deficits $2\pi(1-\alpha_s)$, then the last term
in the
right-hand side of (\ref{a10}) must be replaced by the corresponding
sum over
all $x_s$. If the manifold does not have conical
singularities the last term in (\ref{a10}) vanishes ($\alpha=1$).
Equation
(\ref{a10}) can be also represented in the another
equivalent form which sometimes is more convenient
	\begin{eqnarray}
        W(\tilde{\gamma})-W(\gamma)
		&=& { 1\over 48 \pi } \int_{{\cal M}_{\alpha}} d^2 x
\
	\sigma \left( \tilde{\gamma}^{1/2} \tilde{R} + \gamma^{1/2} R
\right)
\nonumber  \\
	&+&  { 1\over 24 \pi } \int_{\partial {\cal M}_{\alpha}} d x
\
	\sigma \left( \tilde{h}^{1/2} \tilde{k} + h^{1/2} k \right)
\label{W} \\
	&-&  { 1\over 8\pi } \int_{\partial {\cal M}_{\alpha}} d x  \
	\left( \tilde{h}^{1/2} \tilde{k} - h^{1/2} k \right)+{1 \over
12}\left({1
\over \alpha} -\alpha \right)\sigma(x_s)~~~. \nonumber
	\end{eqnarray}
Here
	\begin{eqnarray*}
	 h^{1/2} k  -  \tilde{h}^{1/2} \tilde{k}
	&=& h^{1/2} n^\alpha \partial_\alpha \sigma
	\end{eqnarray*}
and the conformal factor $\sigma$ should be understood as a solution
of the
equation
	\begin{eqnarray*}
	- 2 \gamma^{1/2} \Box \sigma &=& \gamma^{1/2} R -
\tilde{\gamma}^{1/2}
\tilde{R}~~~.
	\end{eqnarray*}

\section{Effective Action and Free Energy of a Scalar Field in Two
Dimensions}
\setcounter{equation}{0}
Let us consider a conformal scalar field $\phi$ on a two-dimensional
manifold.
The  two-dimensional metric is supposed to be independent on the
Euclidean
time. It can be represented in the form
	\begin{eqnarray}
	ds^2 = \exp [ 2\sigma (x) ] \left\{ d \tau^2 + d x^2
\right\}~ \label{mink}
	, \hskip 2cm
	0 \leq \tau \leq \beta ~, \hskip 1cm  x_0 \leq x \leq x_1
{}~~~.
	\end{eqnarray}
The conformal scalar field $\phi$ satisfies the equation
	\begin{eqnarray}
	\triangle \phi = \exp [ - 2 \sigma (x) ] \left\{
{\partial^{2} \over \partial
\tau^2}
	+ {\partial^{2} \over \partial x^2} \right\} \phi = 0
\label{Eq}.
	\end{eqnarray}
For  simplicity we consider the problem with the Dirichlet boundary
conditions
$\phi (x_0) = \phi (x_1) = 0$.

Using the conformal transformation of the effective action   (see
appendix A),
we can reduce the problem of calculation of the effective action on
the
manifold (\ref{mink}) to a calculation of the effective action on a
cylinder
$Q$ with period in Euclidean time $\beta$ and length $L = x_1 - x_0$.
The one-loop effective action on a cylinder $W_1^{Q} ( \beta, L )$
can be
written in the form
	\begin{eqnarray}
	W_1^{Q} ( \beta, L ) &=& {1 \over 2}\ln \mbox{det} (- \mu^2
\triangle)
	= - {1 \over 2} \zeta ' (0)
	+ {1 \over 2} \zeta (0) \ln \mu^2 =
	- {1 \over 2}  \left[ {\partial \over \partial z}
\sum_\lambda
	\left( \mu^2 \lambda \right)^{-z}\right]_{z=0}~~~. \nonumber
	\end{eqnarray}
Here $\mu$ is an arbitrary parameter with a dimensionality of length
and the
generalized $~~\zeta$-function $\zeta(z) = \sum_\lambda [\mu^2
\lambda ]^{-z}$
represents the sum over all eigen values $\lambda$ of the operator $
-
\triangle $. Although the effective action is determined up to the
rescaling of
the parameter $\mu$ all the  physical observables are unambiguously
defined.
For  the Dirichlet boundary conditions the substitution of  the eigen
values
$\lambda_{mn} = ( {2\pi\over\beta})^2 n^2 +({\pi\over L})^2 m^2$ of
the Laplace
operator on the cylinder leads to the relation
	\begin{eqnarray}
	W_1^{Q} ( \beta, L ) &=&
	- {1 \over 2}\left\{ {\partial \over \partial z}
\sum_{m=1}^\infty
\sum_{n=-\infty}^\infty \left[
	\mu^2 \left(
	{4\pi^2\over\beta^2} n^2
	+ {\pi^2\over L^2} m^2
	 \right)
	\right]^{-z}\right\}_{z=0}
	\end{eqnarray}
\begin{eqnarray*}
=-\frac 12\left\{{\partial \over \partial
z}\prod_{m=1}^{\infty}\prod_{n=-\infty}^{\infty}\left({2\pi\mu \over
\beta}n\right)^{-2z}\left(1+{\beta \over 2L^2}{m^2 \over
n^2}\right)^{-z}\right\}_{z=0}~~~.
\end{eqnarray*}
Applying  the formula
	\begin{eqnarray}
	\prod_{n=1}^{\infty} \left( 1 + {a^2 \over n^2} \right) =
{\sinh \pi a \over
\pi a}
	\end{eqnarray}
and representing other infinite sums and products in terms of the
Riemann
$\zeta$-function we  eventually have
	\begin{eqnarray}
	W_1^{Q} ( \beta, L ) &=& \beta {\cal F}  - {\pi \beta \over
24 L}~~~,
	\end{eqnarray}
where
	\begin{eqnarray}
	\beta {\cal F}  =  \sum_{n=1}^\infty
	\ln \left( 1- \exp \left[{-\beta  {\pi  \over L} n}  \right]
\right)~~~.
\label{B7}
	\end{eqnarray}
We demonstrate now that   ${\cal F} $  coincides with the
thermodynamical free
energy  of a  gas of scalar particles in the volume $L$. In
statistical
mechanics the free energy ${\cal F}$ of a quantum system is defined
by a
relation
	\begin{eqnarray}
	\exp [ -\beta {\cal F} ] = \mbox{Tr}  \exp [ - \beta
:\hat{H}: ]~~~.
	\end{eqnarray}
If we choose the basis functions to be eigen-functions of the
Hamiltonian
$\hat{H}=\sqrt{-\partial^2_x}$, the free energy can be expressed in
terms of a
sum over  all dynamical degrees of freedom
	\begin{eqnarray}
	 \beta {\cal F} =  \sum_{n}
	 \ln \left( 1- e^{-\beta \omega_n} \right),
\label{F}
	\end{eqnarray}
where $\beta$ is an inverse temperature, $\omega_n$ are the energy
levels of
the quantum system. Thus we are to know only the spectrum of  the
system to
calculate the free energy.  One can easily solve the Eq.(\ref{Eq})
and find the
energy levels of the system
	\begin{eqnarray*}
	\omega_n &=& {\pi  \over L} n,  \hskip 2cm L = x_1 - x_0
	\end{eqnarray*}
Note that  the mode with $n=0$ should be eliminated from the
summation in
Eq.(\ref{F}), since its amplitude is fixed by the Dirichlet boundary
conditions
and, hence, it is not normalizable and is not a dynamical degree of
freedom.
(For the Neumann boundary conditions  zero modes will contribute to
the free
energy.)

Thus for the Dirichlet boundary conditions the free energy ${\cal F}$
reads
	\begin{eqnarray*}
	 {\cal F}  = {1 \over \beta}  \sum_{n=1}^\infty
	 \ln \left( 1- \exp \left[{-\beta  {\pi  \over L} n}  \right]
\right)~~~,
	\end{eqnarray*}
which coincides with Eq.(\ref{B7}).

Now let us calculate ${\cal F}$ in the high temperature limit, i. e.
when the
length of the cylinder $L$ is much larger that its perimeter $\beta$.
In this limit the distance between the levels is less than
temperature  ${\pi
\over L} \ll {1 \over \beta}$ and the sum over $n$ can be estimated
using the
Euler-McLourain  formula 	\begin{eqnarray*}
	 \sum_{n=1}^\infty f(n) &=& \int _0^\infty d x f(x)
	-\int _0^1 d x f(x) + {1 \over 2} f(1) + \sum_{k=1}^\infty
c_k f ^{(k)}
(1)~~~.
	\end{eqnarray*}
Here the coefficients $c_k$ can be expressed in terms of Bernouli
numbers
	\begin{eqnarray*}
	c_k = (-1)^k {B_{k+1} \over (k+1) ! }
	\end{eqnarray*}
and  the function $f(x)$ is supposed to decrease at infinity together
with all
its derivatives. Substituting here the function  $f(x) = \ln [1-\exp
(-s  x)]$
and taking into account  the relation
	\begin{eqnarray*}
	\ln \Gamma (z) &=& (z-{1 \over 2}) \ln (z) - z + {1 \over 2}
\ln (2 \pi) +
	\sum_{m=1}^\infty {B_{2 m} \over (2 m ) (2 m -1) z^{2 m -1}},
\\
	\mid \mbox{arg} \ z \mid &< &\pi
	\end{eqnarray*}
one can prove that
	\begin{eqnarray}
	    \sum_{n=1}^\infty \ln \left( 1- \exp \left[{-s   n}
\right] \right) =
	-  {\pi^2 \over 6 s } - {1 \over 2} \ln \left[ {s  \over 2
\pi} \right]
	+ {1 \over 24} s + o(s)~~~.
	\end{eqnarray}
For the free energy it leads to a formula
	\begin{eqnarray}
	\beta {\cal F}  = - {\pi L \over 6 \beta} - {1 \over 2 }
	\ln {\beta \over 2 L }  +  {\pi \beta \over 24 \ L} +
o({\beta  \over L})~~~,
	\end{eqnarray}
and hence the effective action reads
	\begin{eqnarray}
	W_1^{Q} = - {\pi L \over 6 \beta} - {1 \over 2 }
	\ln {\beta \over 2 L }  + o({\beta  \over L})~~~.
	\end{eqnarray}
It can be shown that $o(\beta / L)$ is nonanalytical in its argument
and tends
to zero extremely fast when $\beta \leq L$.

Note that by construction $\beta {\cal F}$ for a conformal fields is
conformally
invariant, since the spectrum is conformally invariant.  This
property
distinguishes it from an Euclidean effective  action $W_1$ which
transforms
inhomogeneously under the conformal transformations because of the
conformal
anomaly. Note that the renormalized effective action $W_1^Q(\beta,L)$
and
$\beta {\cal F}$ differ only by the term linear in $\beta~$
\cite{DoKe:78,Alle:86}.

\section{Casimir effect and field fluctuations near the
brick-wall boundary}
\setcounter{equation}{0}
In this Appendix we present a more detail discussion of the field
fluctuations
on the boundary near the horizon and their relation with
the Casimir effect which inevitably arises in the brick-wall
approach. Instead
of the black hole background we consider the quantum field in the
Rindler space
at the inverse temperature $2\pi\alpha$ measured at the point $x=1$,
we put
$\mu=1$ . This simplification is justified by the fact that we
are interested in the effects which happen very close to the horizon
where the
space is similar to a cone (\ref{5.1}).

Assume that the brick wall is at the point $x=\epsilon$ in
coordinates
(\ref{5.1}). The brick wall effective action in this case is the
action on the
part $K_{\alpha,\epsilon}$ of the cone
$C_\alpha$, see Fig. 7.

\begin{figure}
\label{f7}
\let\picnaturalsize=N
\def\picsize{8cm}
\def\picfilename{fig7.eps}
\ifx\nopictures Y\else{\ifx\epsfloaded Y\else\input epsf \fi
\let\epsfloaded=Y
\centerline{\ifx\picnaturalsize N\epsfxsize \picsize\fi
\epsfbox{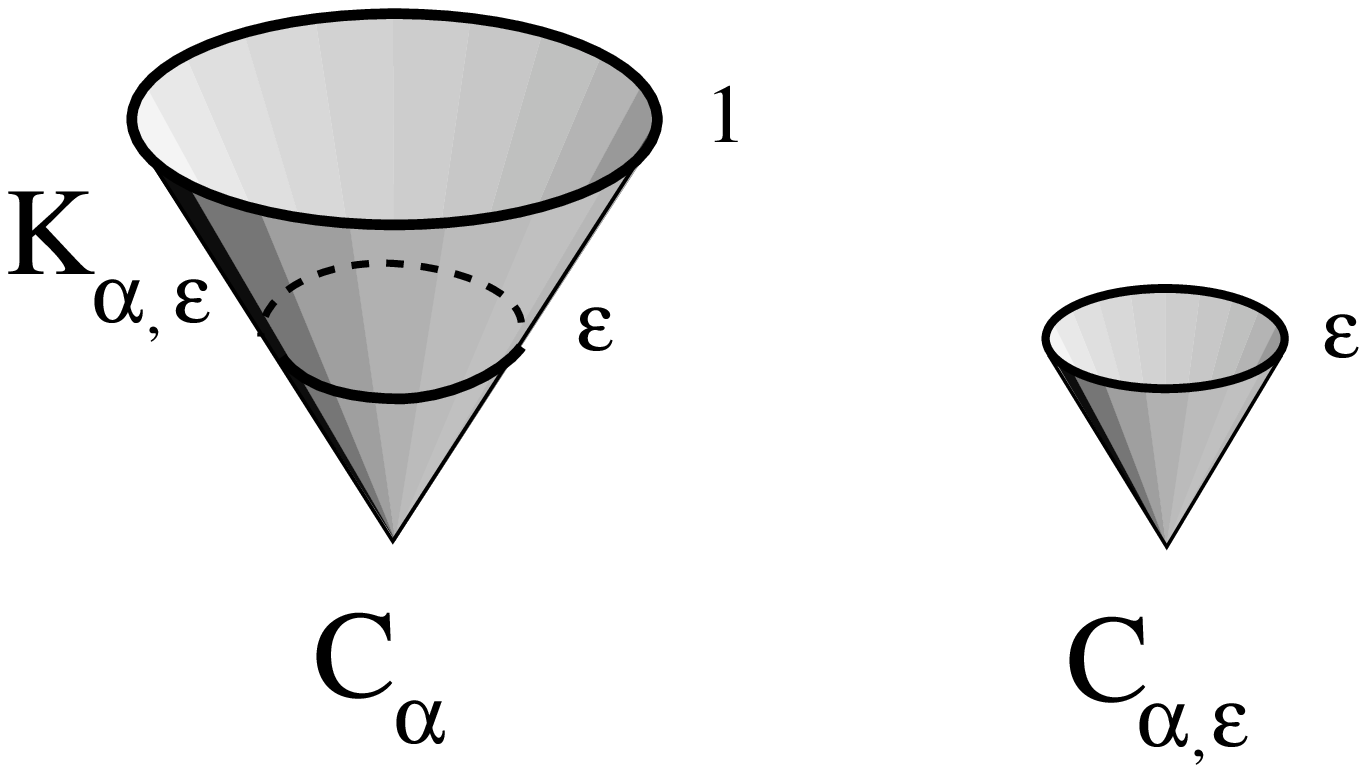}}}\fi
\caption[f7]{Cones}
\end{figure}

Then, as follows from (\ref{4.6}), (\ref{4.7}) and (\ref{24}), the
analog of
the Eq. (\ref{9.13}) for the cone
\begin{equation}\label{C1}
W_1^{BW}(\alpha,\epsilon)=W_1[K_{\alpha,\epsilon}]
=W_1[C_{\alpha}]-W_1[C_{\alpha,\epsilon}]-
W_1^{CAS}(2\pi\alpha,\alpha,\epsilon)~~~,
\end{equation}
\[
W_1^{CAS}(2\pi\alpha,\alpha,\epsilon)=\frac 12\ln{\pi\alpha \over
\ln\epsilon^{-1}}~~~.
\]
Our aim now is to understand how the presence of the Casimir term
$W_1^{CAS}(2\pi\alpha,\alpha,\epsilon)$ is related with the quantum
fluctuations near the point $x=\epsilon$.
This can be done by analysing the path integral representation for
the
partition function on $C_{\alpha}$
\begin{equation}\label{C12}
Z_1[C_\alpha]=
e^{-W_1[C_{ \alpha}]}
=\int \left[ D\phi \right] e^{-I[\phi]}=\int \left[ D\phi \right]
\exp\left(-\frac12 \int \phi_{,\mu}\phi^{,\mu}\right)~~~.
\end{equation}
Here one can divide  the variables into three groups
\begin{equation}\label{13}
Z_1[C_{\alpha}]=\int \left[ D\phi_1 \right] \left[ D\psi
\right]\left[ D\phi_2
\right]e^{-I[\phi]}~~~
\end{equation}
where  $\phi_1$ and $\phi_2$ are the fields in the domain
$x<\epsilon$ and
$x>\epsilon$ respectively,  and  $~\psi =\phi(x=\epsilon)$. In the
each of the
regions one can change  the fields as
\begin{equation}\label{14}
\phi_k=\phi_k '+ \chi_k~~~,
\end{equation}
\begin{equation}\label{15}
\bigtriangleup\chi_k=0~~,~~\chi_k(x=\epsilon)=\psi~~,
{}~~k=1,2~~,~~\chi_2(x=1)=0~~~.
\end{equation}
The new variables $\phi_k '$  satisfy the Dirichlet conditions on the
boundaries of
their domains. Using this fact and that the fields $\chi_k$ are
harmonic, one
can  represent  the classical action in the following way
\begin{equation}\label{16}
I[\phi_1+\phi_2]=I[\phi_1']+I[\phi_2'] +{\cal W}[\psi]
\end{equation}
where  ${\cal W}[\psi]=I[\chi_1]+I[\chi_2]$ for
$\chi_1(x=\epsilon)=\chi_2(x=\epsilon)=\psi$.
The partition function (\ref{13}) is represented now in the form
where
contributions from
the fields $\phi_1$, $\phi_2$ and $\chi$ are completely factorized
\[
Z_1[C_{\alpha}]=\int \left[ D\phi_1' \right]e^{-I[\phi_1']} \int
\left[ D\psi
\right]e^{-\int {\cal W}[\psi]}\int \left[ D\phi_2'
\right]e^{-I[\phi_2']}
\]
\begin{equation}\label{17}
=Z[C_{\alpha,\epsilon}]Z[K_{\alpha,\epsilon}]\int \left[ D\psi
\right]e^{-\int
{\cal W}[\psi]}~~~.
\end{equation}
The first multiplier in (\ref{17}) is the partition function on a
cone of the
small radius $\epsilon$,
the second one is the partition function on the space
$K_{\alpha,\epsilon}$,
which is determined by the brick wall action
$W_1^{BW}(\alpha,\epsilon)$
\begin{equation}\label{C17}
Z[K_{\alpha,\epsilon}]=e^{-W_1[K_{\alpha,\epsilon}]}=
e^{-W_1^{BW}(\alpha,\epsilon)}~~~.
\end{equation}

The left integral over $\psi$ describes the quantum fluctuations of
the field
in the point $x=\epsilon$. Let us show that it reproduces explicitly
the
Casimir term in the effective action. Indeed, Eqs. (\ref{15}) have
the
following solutions \begin{equation}\label{18}
\chi_1(x,\tau)=\sqrt{1 \over
\pi\alpha}\sum_{n=1}^{\infty}\left(\psi^{(1)}_n\cos{ n \tau \over
\alpha}+
\psi^{(2)}_n\sin{ n \tau\over \alpha }\right)\left({x \over
\epsilon}\right)^{n
\over\alpha }~~~,
\end{equation}
\begin{equation}\label{19}
\chi_2(x,\tau)=\sqrt{1 \over \pi\alpha}
\sum_{n=1}^{\infty}\left(\psi^{(1)}_n\cos{ n \tau \over \alpha}+
\psi^{(2)}_n\sin{ n \tau \over \alpha}\right)
\left({\epsilon \over x }\right)^{n \over \alpha }{1 -x ^{2 n \over
\alpha}
\over  1 -\epsilon ^{2 n \over \alpha}}
+{\psi_0 \over \sqrt{2\pi\alpha}}\ln{x/\epsilon}~~~
\end{equation}
where $\psi^{(k)}_n,~\psi_0$ are the Fourier coefficients of the
field $\psi$
on the boundary:
\begin{equation}\label{20}
\psi(\tau)=\sqrt{1 \over
\pi\alpha}\sum_{n=1}^{\infty}\left(\psi^{(1)}_n\cos{ n
\tau \over \alpha}+
\psi^{(2)}_n\sin{ n \tau \over \alpha}\right)+{\psi_0 \over
\sqrt{2\pi\alpha}}~~~
\end{equation}
defined with respect to the orthonormal basis on the circle
$0\leq\tau\leq
2\pi\alpha$.
This gives the action up to the terms of $O(\epsilon)$ in the form
\begin{equation}\label{21}
{\cal W}[\psi]=I[\chi_1]+I[\chi_2]={1 \over
\alpha}\sum_{n=1}^{\infty}\left[
(\psi^{(1)}_n)^2+
(\psi^{(2)}_n)^2\right] + \left(2\ln{1 \over
\epsilon}\right)^{-1}\psi_0^2+
O(\epsilon)~~~.
\end{equation}
The integral over $\psi$ has the Gaussian form and can be evaluated
exactly.
The integration measure can be written up to a normalization
numerical
coefficients as
\begin{equation}\label{measure}
\left[D\psi\right]=\epsilon^{1/2}
d\psi_0\prod_{n=1}^{\infty}\epsilon^{1/2}
d\psi_n^{(1)}\prod_{n=1}^{\infty}\epsilon^{1/2} d\psi_n^{(1)}~~~,
\end{equation}
where the multiplier $\epsilon^{1/2}$ is the heritage of the
definition of
the covariant measure which includes the factor ${g}^{1/4}$ at
$x=\epsilon$.
Thus, the result of the integration over fields $\psi$
looks as
\begin{equation}\label{28}
\int \left[ D\psi \right]e^{-\int {\cal W}[\psi]}= {\cal N}
\left(\epsilon\ln{1 \over \epsilon}\right)^{\frac 12}\exp\left(
\sum_{n=1}^{\infty}\ln (\alpha \epsilon)\right)
\end{equation}
(${\cal N}$ is the numerical constant) which after regularization of
the
infinite sum with the help of the Riemann zeta-function $\zeta_R(z)$
\[
\sum_{n=1}^{\infty}=\lim_{z\rightarrow 0}\sum_{n=1}^{\infty}n^{-z}=
\zeta_R(0)=-\frac 12
\]
gives the Casimir term
\begin{equation}\label{29}
\int \left[ D\psi \right]e^{-\int {\cal W}[\psi]}={\cal N}
\left({\ln{\epsilon^{-1}} \over \alpha}\right)^{\frac 12}=
{\cal N}e^{W_1^{CAS}(\alpha,\epsilon)}~~~.
\end{equation}
The Eqs. (\ref{17}) and (\ref{29}) result in the formula
\begin{equation}\label{121}
e^{-W_1[C_{\alpha}]}=Z_1[C_{\alpha,\epsilon}]Z[K_{\alpha,\epsilon}]
e^{W_1^{CAS}(\alpha,\epsilon)}=e^{-(W[C_{\alpha,\epsilon}]+
W[K_{\alpha,\epsilon}]-W_1^{CAS}(\alpha,\epsilon))}
\end{equation}
which, obviously, reproduces the relation (\ref{C1}) between the
brick wall
the action $W_1^{BW}$ and action $W_1[C_\alpha]$ on the the cone,
which we
obtained
earlier by the conformal transformation.


\begin{thebibliography}{000}
\bibitem{fn1} A similar off-shell procedure was used
in\cite{Teitelboim:94,CaTe:95,Teitelboim:95} to give an
interpretation of the
Bekenstein-Hawking entropy. The idea is to link the entropy with the
surface
degrees of freedom which give rise to  conical singularities for the
off-shell
black-hole geometries.  This  possible
application of the off-shell method is not discussed in the present
paper.
\bibitem{fn2} This approach differs essentially from that used in
Refs.\cite{SU,DLM}, where the calculations begin with the bare
classical action
while the renormalization of coupling constants and restoration of
the
observable values of the parameters is performed only at the end.
\bibitem{fn3} Throughout the paper we use the subscripts $0$ and $1$
to
distinguish tree-level and one-loop quantities, respectively. The
superscripts
are used to distinguish between the quantities calculated by
different methods
or in different approaches.
\bibitem{fn4} Strictly speaking, in the presence of the ultraviolet
divergences
the evaluation of the effective action (Eqs.(\ref{1.2}) -
(\ref{1.4})) needs a
modification. One must first decompose the bare classical action $I$
into the
piece $I_{ren}$ expressed through the renormalized coupling constants
and a
divergent part $I_{div}$ which is of the next order in the Planck
constant
$\hbar$ as compared to $I_{ren}$.
The stationary point approximation is considered with the respect to
$I_{ren}$,
so that the classical solution $\phi_0$ is expressed in terms
of the renormalized constants. On the other hand, $I_{div}$
in the combination with $W_1$ compensates the divergences in this
functional
and
produces the renormalized one-loop correction to the classical
action.
\bibitem{fn5} Note that parameters $\mu$ which enter the
Eq.(\ref{2.7}) and
Eq.(\ref{2.71a}) are logically different. For simplicity we identify
them, so
that $C$ in Eq.(\ref{2.75a}) is now independent from $\mu$ and,
hence, it is
just a numerical constant.
\bibitem{fn6} There is an elegant reformulation of the 't Hooft brick
wall
model. The idea\cite{DLM} is to use a Pauli-Villars regularization
and to
introduce a number of fictious fields with very large masses $\sim
m_{PV}$. The
role of these fields is to cancel contributions of very high energy
modes into
the free energy and entropy of the black hole. It makes the latter
quantities
finite in the region near the horizon without introduction of any
brick walls.
As a consequence, this regularization results in the divergences
$\sim \ln
m_{PV}$ when
$m_{PV}\rightarrow\infty$. So in this approach the parameter
$m_{PV}^{-1}$
replaces the brick wall parameter $\epsilon$.
\bibitem{fn7} In a static space $M_B^{\alpha}$ having the conical
singularity
at $r=r_h$ with the deficit angle $2\pi(1-\alpha)$  the integral of
the
curvature
with a test function $f(r)$ can be represented as \cite{FS1}
\[
\int_{M_B^{\alpha}}f(r)R(r)\gamma^{1/2}d^2x=\alpha
\int_{M_B}f(r)R(r)\gamma^{1/2}d^2x+4\pi(1-\alpha)f(r_h)~~~.
\]
We use this relation for $f(r)=r^2$, $r_h=r_+$. The first term in the
right
hand side  is calculated on the regular manifold
$M_B=\left.M_B^\alpha\right|_{\alpha=1}$, while the second term  is
present
because of  the conical singularity.
\end{thebibliography}
\end{document}